\documentclass{article}
\usepackage[utf8]{inputenc} 

\RequirePackage[T1]{fontenc}
\RequirePackage[tt=false, type1=true]{libertine}

\usepackage{hyperref}       
\usepackage{url}            
\usepackage{booktabs}       
\usepackage{amsfonts}       
\usepackage{nicefrac}       
\usepackage{microtype}      
\usepackage{xcolor}         
\usepackage{paper}
\usepackage{adjustbox}
\usepackage{wrapfig}
\usepackage{xspace}
\usepackage{cprotect}

\usepackage{makecell}
\usepackage{tcolorbox}

\newtoggle{arxiv}
\toggletrue{arxiv}

\iftoggle{arxiv}
{
\setlength{\textwidth}{6.5in}
\setlength{\textheight}{9in}
\setlength{\oddsidemargin}{0in}
\setlength{\evensidemargin}{0in}
\setlength{\topmargin}{-0.5in}
\newlength{\defbaselineskip}
\setlength{\defbaselineskip}{\baselineskip}
\setlength{\marginparwidth}{0.8in}
}

\usepackage{courier}
\usepackage{listings}
\definecolor{mygreen}{rgb}{0,0.6,0}
\definecolor{mygray}{rgb}{0.5,0.5,0.5}
\definecolor{mymauve}{rgb}{0.58,0,0.82}

\title{\bf \textsf{Multi-Relational Analytics and Its Applications to Data Insights}}

\DeclareMathOperator{\region}{{\sf Region}}

\DeclareMathOperator{\SliceSelect}{{\sf SliceSelect}}

\DeclareMathOperator*{\represent}{{\sf Represent}}

\DeclareMathOperator*{\flatten}{{\sf Flatten}}

\makeatletter
\DeclareRobustCommand*\cal{\@fontswitch\relax\mathcal}
\makeatother

\hypersetup{
  colorlinks,
  linkcolor={red},
  citecolor={green},
  urlcolor={magenta}
}

  {]\par} 

  {]\par} 

  {]\par} 

\title{Multi-Relational Algebra for Multi-Granular Data Analytics}
\author{
  Xi Wu,
  Eugene Wu\textsuperscript{\textdagger},
  Zichen Zhu,
  Fengan Li\textsuperscript{\textdaggerdbl}, 
  Jeffrey F. Naughton\textsuperscript{\textdaggerdbl}
   \\  {\small Google, \textsuperscript{\textdagger}Columbia University,
  \textsuperscript{\textdaggerdbl}Celonis}
}
\date{}

\begin{document}

\maketitle

\begin{abstract}

In modern data analytics, analysts frequently face the challenge of searching for desirable \emph{entities} by evaluating, for each entity, a collection of its
\emph{feature relations} to derive key analytical properties. This search is challenging because the definitions of both entities and their
feature relations may span multiple, varying granularities. Existing constructs such as {\sf GROUP BY CUBE}, {\sf GROUP BY GROUPING SETS},  {\sf ARRAY AGGREGATE}, WINDOW functions, OLAP cube, and various data explanation paradigms aim to facilitate such analyses, but all exhibit limitations in terms of composability,  clear specifications, and performance. To address these challenges, we introduce Multi-Relational Algebra (MRA), which generalizes relational algebra with two core data abstractions: \textbf{RelationSpace}, for managing collections of relations, and \textbf{SliceRelation}, which structures data around entities with corresponding relation-valued features. MRA introduces a rich set of operators for transforming data between these representations, enabling complex multi-granular analysis in a modular and declarative way. An early version of MRA is in production at Google, supporting diverse data insight applications. This paper describes the motivation for MRA, its formalism, implementation, and future opportunities.

\end{abstract}

\section{Introduction}
\label{sec:intro}

In modern data analytics, analysts are often tasked with identifying interesting 
entities---such as customer segments or user groups---for applications like 
targeted marketing or system debugging. 
A key challenge is that the precise definitions of these entities and their most 
salient features are typically not known \emph{a priori}. 
Consequently, analysts must iteratively explore a vast definition space of 
granularities and metrics while simultaneously analyzing the data. 
For instance, consider an analyst examining ad performance using a relation 
with the schema \textsf{[Device, Browser, Date, Cost, Clicks]}. 
An \emph{entity} might be defined at the granularity of a single device
or a device-browser combination. 
Critically, the \emph{features} used to evaluate these entities are also 
multi-granular and often more complex than simple scalar aggregates. 
A feature might itself be a relation, such as a cost-per-click (Cpc) time series 
(\textsf{Cpc = SUM(Cost)/SUM(Clicks)}), which can then serve as input to a 
custom statistical model to detect outliers like ``sudden spikes.''

Unfortunately, this analytical pattern, which involves reasoning over collections 
of relations and composing arbitrary user-defined functions, exposes the fundamental 
limitations of existing analytical paradigms. 
On one hand, OLAP systems, queried via languages like MDX, are designed for 
multi-granular \emph{manual exploration}. However, their logical model is the \emph{multidimensional 
array}, not the relation. This makes them adept at navigating pre-aggregated scalar 
values but ill-equipped to handle the relation-valued features or complex statistical 
transformations central to our problem. Furthermore, their non-relational nature 
hinders composability within the broader SQL ecosystem. 
On the other hand, SQL is built on a relational foundation that is also a poor fit. 
The core mismatch is that classical relational algebra operates on individual 
relations, whereas this problem is conceptually about operations over \emph{sets of relations}. 
To bridge this gap, practitioners resort to cumbersome workarounds. For example, they use 
extensions like \textsf{GROUP BY CUBE}, which collapses many logical groupings into a 
single physical table, using \textsf{NULL} padding to represent the different levels of aggregation. 
While this computes the necessary data, it obfuscates the logical structure of the 
groupings, forcing analysts to write complex procedural queries to decode and handle 
each granularity separately. This approach creates significant \emph{logical-physical entanglement}, 
diverging sharply from the declarative ideal of SQL.

The difficulty of implementing these workloads in SQL-based systems, such as 
GoogleSQL, BigQuery, and F1 Query~\cite{F1Query}, has led many teams at Google to 
abandon the relational ecosystem entirely. Instead, they adopt general-purpose 
data processing frameworks like Apache Beam~\cite{PythonBeam}. Beam's flexible 
\textsf{PCollection} model, which represents data as arbitrary key-value pairs, 
makes it easier to model varying entities and features and to integrate with 
Python's rich statistical libraries. 
However, this shift is undesirable: it introduces system fragmentation, creates 
interoperability challenges, and forgoes decades of performance optimizations built 
into modern relational engines.

To address these challenges in a principled way, we present \textbf{Multi-Relational 
Algebra (MRA)}, an extension of classical relational algebra designed specifically 
for multi-granular analytics. MRA's design centers on two new data abstractions. 
The first, \textsf{RelationSpace}, is inspired by OLAP cubes and provides a formal 
mechanism for managing and querying a collection of relations defined at different 
granularities. 
The second, \textsf{SliceRelation}, is inspired by the nested relational model and 
restructures data to align entities with their corresponding features. 
Each \emph{slice tuple} in a \textsf{SliceRelation} is keyed by an entity (called a 
\emph{region}) and contains a set of relation-valued features. 
MRA provides the expressive power needed for complex feature analysis, while
remaining fully compatible with the classical relational model. 
It intentionally constrains nesting to a single, structured level---a design choice 
that simplifies the user's mental model and enables a rich set of specialized operators 
that would be difficult to define for arbitrary nesting.
Complementing these data types, MRA introduces a rich set of novel operators to 
transform them. 
Together, these constructs enable analysts to compactly represent and compute over 
entire sets of relations for multi-granular analytics. 
Table~\ref{tab:data-objects-of-different-paradigms} and 
Table~\ref{tab:operators-of-different-paradigms} compare these data objects and 
operators against existing paradigms, respectively.

We have implemented MRA as a domain-specific language (DSL) exposed via 
Table-Valued Functions (TVFs) and deployed it as a production service at Google. 
Its adoption has been broad and impactful: between April and June 2025, the service 
was used by over 600 teams, handling an average of 10 million queries per day and 
enabling insights that have led to significant revenue gains. 
Notably, several teams have migrated complex analytical workloads from Apache Beam to MRA, 
benefiting from improved performance and deeper integration with Google's relational data ecosystem.

\begin{table}[htbp]
\centering
\renewcommand{\arraystretch}{1.4}
\begin{tabular}{p{3cm}||p{12cm}}
\hline
\textbf{Data Type} & \textbf{Representation} \\
\hline
Relation &
\{ ($\mathsf{a_1}$=$\mathsf{v_1}$, $\mathsf{a_2}$=$\mathsf{v_2}$, \ldots), \ldots \}; a set of tuples. \\
\hline
Multidimensional Array (MDX Cube) &
$[\, [\ldots],\ [\ldots],\ [\ldots],\ \ldots\,]$; each $[\ldots]$ is a potentially nested array. \\
\hline
PCollection (Beam) &
$\{\, (\mathsf{k_1}, \mathsf{v_1}),\ (\mathsf{k_2}, \mathsf{v_2})\ \ldots\,\}$; a set of KV pairs, where K and V can be arbitrary objects. \\
\hline
MRA RelationSpace &
$\{\, \mathsf{R_1},\ \mathsf{R_2},\ \ldots\ \mid\ \mathsf{dimensions} = \ldots \,\}$; a set of relations with dimensions. \\
\hline
MRA SliceRelation &
\{ $\mathsf{region_1}$:\ \ ($\mathsf{feature\_schema_1}$=$\mathsf{R_1}$, $\mathsf{feature\_schema_2}$=$\mathsf{R_2}$, \ldots ), \ldots | $\mathsf{dimensions}$=$\ldots$ \};
a set of slice tuples with dimensions, where each slice tuple is keyed by a region tuple, and the value has a tuple structure with a set of feature columns.\\
\hline
\end{tabular}
\caption{Data types and their representations.}
\label{tab:data-objects-of-different-paradigms}
\end{table}

\begin{table}[htbp]
\centering
\renewcommand{\arraystretch}{1.4}
\begin{tabular}{l||p{12cm}}
\hline
\textbf{Paradigm} & \textbf{Operator Semantics} \\
\hline
Relational Algebra &
$\mathsf{Relation} \rightarrow \mathsf{Relation}$. Mostly closed under $\mathsf{Relation}$, except for joins which merge multiple relations into one. \\
\hline
Apache Beam &
$\mathsf{PCollection} \rightarrow \mathsf{PCollection}$. Each $\mathsf{PCollection}$ is a set of arbitrary $(k,v)$ pairs, allowing for arbitrarily nested key and value structures. \\
\hline
OLAP Cube (MDX) &
$\mathsf{Multidimensional\ Array\ (Cube)} \rightarrow \mathsf{Cellset}$. A multi-type transformation: the result is not another cube but a $\mathsf{Cellset}$, which affects composability in an algebraic context. \\
\hline
MRA &
\parbox[t]{12cm}{
MRA has three data types: $\mathsf{Relation}$, $\mathsf{RelationSpace}$, and $\mathsf{SliceRelation}$. The core algebra includes: \\
- $\mathsf{Relation} \rightarrow \mathsf{RelationSpace}$ (lift with dimensions) \\
- $\mathsf{RelationSpace} \rightarrow \mathsf{SliceRelation}$ (Represent) \\
- $\mathsf{SliceRelation} \rightarrow \mathsf{SliceRelation}$ (MRA transformations) \\
- $\mathsf{SliceRelation} \rightarrow \mathsf{RelationSpace}$ (Flatten)
} \\
\hline
\end{tabular}
\caption{Operator semantics across different data analysis paradigms.}
\label{tab:operators-of-different-paradigms}
\end{table}

In the rest of the paper we give more details on the use cases and motivations, describe the core algebra, and discuss our initial implementation at Google. The content is organized as follows:
\begin{itemize}
    \item In Section~\ref{sec:mga} we more present more details of the use cases and motivations.
    \item In Section~\ref{sec:data-model} we present the data model of MRA, extending the data model of RA.
    \item Section~\ref{sec:operations-sr} and Section~\ref{sec:operations-rs} present algebraic operations.
    \item We describe our initial implementation at Google in Section~\ref{sec:impl} and its applications in Section~\ref{sec:applications}.
    \item Finally, we discuss related work in Section~\ref{sec:related}, and conclude in Section~\ref{sec:conclusion}.
\end{itemize}


\section{Multi-Granular Analytics and Multi-Relational Algebra}
\label{sec:mga}
In this section we describe use cases that lead us to the Multi-Relational Algebra, and an overview of how MRA works for these use cases.
To start with, in this paper we model \emph{entities} as \emph{tuples in the classic relational model}. For example, an individual device entity is modeled as a tuple such as (Device=Pixel), an individual browser entity is modeled as a tuple such as (Browser=Chrome), and a device-browser entity is modeled as a tuple such as (Device=Pixel, Browser=Chrome). A \emph{feature} is modeled as \emph{a relation in the classic relational model}. For example we may consider a timeseries feature, say a relation with schema [Date, Cpc], by aggregating the cost-click data with {\sf Cpc=SUM(Cost)/SUM(Clicks)}. By \emph{granularity}, intuitively it means the different attributes that need to participate the aggregation for generating required entities and features. For example, to generate Browser entities and [Date, Cpc] feature, we need to group by [Browser, Date], and to generate Device entities and [Date, Cpc] feature, we in turn need to group by [Device, Date].

For analytics that involve only uniform entity and feature granularity, traditional analytics paradigms, such as SQL and OLAP cube, have handled them very well. Therefore, the interesting cases arise only when we begin to vary entity and feature granularity. To better guide the readers navigate the various problems and solutions, we organize the content below as follows:

\begin{itemize}
    \item We first study tasks where the entities have uniform-granularity, but the features have varying granularity. We show that there are already manifestation of problems at this level.
    \item We then move to consider the case where both entities and features have varying granularity. There are more problems that arise in this case, with problems described for the simpler tasks become more pronounced.
    \item We then present the Multi-Relational Algebra (MRA) way to address all the problems under a coherent algebraic framework.
    \item Finally, we describe more follow-up analyses that are easily enabled by MRA due to its composability.
\end{itemize}

In the rest of the section, we use $X$ to refer to a base relation with schema [Device, Browser, Date, Cost, Clicks].






\subsection{Uniform Entities and Varying Features}
As a simplification, let us first consider the case where entities have a uniform granularity but features have varying granularity. Consider a set of devices (entities) and a set of analytical tasks which are simple variants of each other:
\begin{description}
\item[(T1)] \emph{Find devices that have cost-per-click timeseries anomalies and have total cost > 1000.}
\item[(T2)] \emph{Find devices that have cost-per-click timeseries anomalies and have total cost <= 1000.}
\item[(T3)] \emph{Find devices that have cost timeseries anomalies and have aggregated cost-per-click > 10.}
\item[(T4)] \emph{Find devices that have cost timeseries anomalies and have aggregated cost-per-click <= 10.}
\end{description}
For these, we want to output a table with schema [Device, Date, Cpc, IsAnomaly, TotalCost]. Here, to find out timeseries anomalies, we want to apply a statistical analysis, for example {\sf CausalImpact} analysis~\cite{BGKRS15}, which requires a highly non-trivial statistical estimator.
Just for these seemingly different tasks, we observe different options of using SQL (for simplicity in the following we present options for \textbf{(T1)}, others are similar).

\begin{approach}[\textbf{Option U1}]
\label{approach:sql-1}
We perform the following operations:
\begin{enumerate}
\item Find devices with total cost > 1000 and store the results in a table {\sf tmp1} with schema [Device, TotalCost]. The relational algebra expression is
$$\rm tmp_1 = \sigma_{Cost>1000}(\gamma_{Device, SUM(Cost)}(r))$$

\item Compute a daily-cpc table for different devices, store the results in a table {\sf tmp2} with schema [Device, Date, Cpc]. The relational algebra expression is
$$\rm tmp_2 = \gamma_{Device, Date, SUM(Cost)/SUM(Clicks) \rightarrow Cpc}(r)$$

\item Join the two tables from previous steps and get a table with schema [Device, Date, Cpc, TotalCost]. The relational algebra expression is
$$\rm  tmp_3 = tmp_1 \bowtie_{tmp_1.Device=tmp_2.Device} tmp_2$$

\item Compute array-values of [Date, Cpc] for each device. The relational algebra expression is
$$\rm tmp_4 = \gamma_{Device, ARRAY\_AGG([Date, Cpc]) \rightarrow DateCpcArray}(tmp_3)$$

\item Detect anomalies by applying UDF (User Defined Function) to the array-valued column DateCpcArray. The relational algebra expression is:
$$\rm results = \mu_{CausalImpactAnalysis(DateCpcArray) \rightarrow DateAnomalyArray}(tmp_4)$$
where $\mu$ denotes the operator to mutate some columns by applying functions to the designated columns of each tuple, with all the other columns  intact. 
\end{enumerate}
\end{approach}
Note that, strictly speaking, for the final results we need to unnest the array, though this is minor for our discussion. The same comment applies to options below.

\begin{approach}[\textbf{Option U2}]
\label{approach:sql-2}
We perform the following operations:
\begin{enumerate}
\item Sames as step (1) of {\bf U1}.

\item Sames as step (2) of {\bf U1}.

\item Sames as step (3) of {\bf U1}.

\item Detect anomalies by applying UDA (User Defined Aggregation) to [Date, Cpc] for each device. The relational algebra expression is
$$\rm results = \gamma_{Device, CausalImpactUDA([Date, Cpc]) \rightarrow DateAnomalyArray}(tmp_3)$$
\end{enumerate}
\end{approach}

\begin{approach}[\textbf{Option U3}]
\label{approach:sql-3}
We perform the following operations: (1) and (2) can be performed in parallel,
\begin{enumerate}
\item Same as step (1) of {\bf U1}.

\item First do step (2) of {\bf U1}, then detect timeseries anomalies using either step (4) and (5) of the option {\bf U1} (the array-aggregate+UDF approach), or step (4) of the option {\bf U2} (the UDA approach). The result is {\sf tmp3} with schema [Device, DateAnomalyArray].
    \item Join {\sf tmp1} and {\sf tmp3} on Device to get the final results.
\end{enumerate}
\end{approach}

\begin{approach}[\textbf{Option U4 -- Monolithic UDA}]
\label{example:sql-4}
We can also do a monolithic UDA: group by [Device] and apply an UDA to [Date, Cost, Clicks]: as we go through each tuple of [Date, Cost, Clicks] for a device, we sum up Cost across dates, as well as accumulate these tuples, and finally, we perform the anomaly detection after we get all the tuples for the device. With this approach, we can filter \emph{within} the UDA for devices with cost <= 1000, avoiding costly anomaly detections when possible. The relational algebra expression is:

$$\rm results = \gamma_{Device, TotalCostAndCausalImpactUDA([Date, Cost, Clicks]) \rightarrow DateAnomalyArrayWithTotalCostFiltered}(r)$$
\end{approach}

We observe two categories of problems:

\textbf{Logical-Physical Entanglement}. For a single, conceptually simple, analytical task, an analyst has to choose from several different and complex SQL queries. The best choice depends on physical implementation details and data size, not just the logical goal of the analysis. Specifically, for small data \textbf{U3} may be the best due to parallel processing. However, once data gets large then it is preferable to choose \textbf{U1} or \textbf{U2}, since the statistical analysis such as {\sf CausalImpact} can be costly, and we want to first filter the devices that we want to perform analysis. To this end, \textbf{U4} avoids join but it breaks the abstractions by squeezing a {\sf SUM} aggregation into user-defined aggregation. To this end, we note similar observations have been made in recent work~\cite{deshpande2025beyond} that the classic relational abstractions are low-level and closely tied to the physical layout of the data.

\textbf{Inefficient Execution and Data Handling}.   There are two dominant sources of inefficiency.
 1) There is considerable overhead in converting data formats.  Using UDAs forces a costly conversion of data from a columnar format to a row-based format. The process is even worse for UDFs, which can require multiple data format conversions.
 2) Join processing is inefficient.   Joins are a form of "row-by-row processing" , which is inefficient when the logic of the analysis only requires a single join per entity.

\subsection{Varying Entities and Varying Features}

Things get worse if we consider varying entity-granularity. For example, now analysts are interested in individual browsers and also different device-browser combinations. Then we have again different options to consider.

\begin{approach}[\textbf{SQL for Varying Entity-Granularity: Option V1}]
\label{approach:varying-entity-sql1}
We perform the following operations:
\begin{enumerate}
\item We compute different groupings: [Device, Cost, Cpc], [Device, Date, Cost, Cpc], [Browser, Cost, Cpc], [Browser, Date, Cost, Cpc], [Device, Browser, Cost, Cpc], [Device, Browser, Date, Cost, Cpc], and store them in six different tables (so we need table names).

\item Then for each entity-granularity, perform analysis using one of {\bf (U1) -- (U4)}.  For instance, for browser entities, we can apply one of {\bf (U1) -- (U4)} to the tables [Browser, Cost, Cpc] and [Browser, Date, Cost, Cpc].
\end{enumerate}
\end{approach}

\begin{approach}[\textbf{SQL for Varying Entity-Granularity: Option V2}]
\label{approach:varying-entity-sql2}
Instead of preparing individual relations, we can use {\sf GROUP BY CUBE}. Specifically:
\begin{enumerate}
\item We use {\sf GROUP BY CUBE} or {\sf GROUP BY GROUPING SETS} to prepare the aggregated data as one table. For example:
\begin{lstlisting}[language=SQL,mathescape=true,label={lst:groupby-cube}]
SELECT
  Device, Date, SUM(Cost) AS Cost, SUM(Cost)/SUM(Clicks) AS Cpc
FROM
  X
GROUP BY CUBE
  Device, Date
\end{lstlisting}

\item Then we modify any one of {\bf (U1) -- (U4)} to analyze the cube results from (1).
\end{enumerate}
\end{approach}

Clearly, we inherit the same row-oriented data format conversion and join processing from the uniform-entity case, but now more problems arise.

\textbf{Inflexible Data Modeling and Composability}.

\begin{itemize}
\item \textbf{Cumbersome Management of Logical Data}. Analysts are forced to manage a large number of different tables, with the number being a product of the different entity and feature granularities.

\item \textbf{Broken Composability}. The ability to compose new analyses from the results of prior ones is broken. Users must either manually manage many tables or work with the cumbersome results from {\sf GROUP BY CUBE}.

\item \textbf{Forced Materialization}. Using {\sf GROUP BY CUBE} forces the materialization of all data groupings. It is not possible to materialize only a specific group of interest (e.g., only device-level data) because the logical abstraction is at the table granularity, not a group of rows within a table.
\end{itemize}

\textbf{Logical-Physical Entanglement}.
\begin{itemize}
\item \textbf{Entanglement Compounded by Inflexible Modeling.} Using {\sf GROUP BY CUBE} to manage varying granularities exacerbates the entanglement: \textbf{(1) Schema information loss}. The relational algebra forces us to logically output one table with {\sf NULL/ALL} padding and so there is a \emph{schema information loss}: we can no longer use the schema to distinguish different groupings, and instead we use a value padding approach, and rely on padded values, such as {\sf NULL/ALL}, or grouping id, to record grouping structure. This makes things harder to work with for the follow-up analysis. \textbf{(2)} \textbf{Logical information redundancies}. The value padding approach {\sf GROUP BY CUBE} needs to tag every tuple, and thus induces \emph{logical information redundancies}. \textbf{(3)} \textbf{Cumbersome processing}. Perhaps the biggest issue is this, even if we have one single table of aggregated data, we still have to analyze \emph{each individual group} independently. For example, suppose we rely on grouping id to identify groups. Then, we need to do group by [Device] with UDA on the data with {\sf grouping\_id=101}, group by [Browser] with UDA on the data with {\sf grouping\_id=011}, and group by [Device, Browser] with UDA on the data with {\sf grouping\_id=111}. The reason for this independent processing is because, for example, directly group by [Device] on the cube results will mix the data in the group with {\sf grouping\_id=101} and the data in the group with {\sf grouping\_id=111}. Note that the number of queries the analyst has to write is the same as the number of groups one needs to examine.

\item \textbf{Manual Optimization Coding}. With varying entity-granularity, there is more performance optimization opportunities. For example, one potential is to exploit cross grouping-set dependencies: If a device entity, say (Device=Pixel), already has cost <= 1000, then we can skip all the entities of the form (Device=Pixel, Browser=*). This is known in the data mining literature as the Apriori property~\cite{AprioriAlgorithm}. There are different ways to apply the Apriori property for pruning (e.g., depth-first search vs. breadth-first search), and one can program these with careful applications of join. However, these further deviate from the logical goal of queries.
\end{itemize}

\subsection{The Multi-Relational Algebra way}

To address the above problems, we introduce two new data types: RelationSpace and SliceRelation.

\textbf{RelationSpace}. We begin with problems related to data management and modeling. These issues are not new and have long motivated the design of OLAP cubes, which adopt a different logical abstraction based on the \emph{multi-dimensional array} model. In this model, attributes are categorized as \emph{dimensions} and \emph{measures}. For example, [Device, Date] may serve as dimensions and [Cost] as a measure. At the base level, the dimensions [Device, Date] define a two-dimensional array where each cell contains the cost value for a given device on a specific date. Crucially, OLAP systems interpret subsets of dimensions as defining different levels of \emph{granularity}. For instance, if Cost is defined using {\sf SUM}, then projecting onto [Device, Cost] yields a one-dimensional array of total cost per device (aggregated over dates), while [Date, Cost] gives cost per date (aggregated over devices), and selecting only [Cost] produces a scalar: the total cost across all devices and dates. These semantics make dimension subsets central to how OLAP queries operate. As a result, the logical interface of OLAP systems—exposed via languages like MDX~\cite{MDX}—departs significantly from SQL, despite superficial similarities. Even though OLAP cubes can be implemented over relational engines (as in ROLAP), their logical model is not directly interoperable with SQL and remains a special-purpose analytics paradigm.

RelationSpace offers a new logical abstraction for managing collections of relations in a way that remains fully compatible with the relational model. Compared to standard relational algebra, it introduces the concept of a \emph{dimension set}, which provides a unified mechanism for identifying and organizing related tables. The core intuition is that, if we store data at different granularities as separate relations, then the combinations of dimensions used in OLAP cube queries correspond naturally to these dimension sets. RelationSpace makes this correspondence explicit, allowing multiple relations to be grouped and queried by their dimensional schema, while preserving interoperability with SQL-based systems.

\begin{definition}[\textbf{RelationSpace}]
\label{def:relation-space}
Let $D$ be a set of attributes called dimension attributes (or simply dimensions), and $V$ a set of attributes called value attributes. A RelationSpace $\Psi$ is a data object that consists of several relations of distinct schemas, $\calF_1, \dots, \calF_k$, where the schema of each relation is a union of a \emph{distinct} subset of $D$ and a subset of $V$. That is, for any $i \neq j$, $\calF_i \cap D \neq \calF_j \cap D$. The schema of $\Psi$, denoted as $\schema(\Psi)$, is defined as the tuple $\langle D, V, \overline{\calF}\rangle$, where we denote $\overline{\calF} = [\calF_1, \dots, \calF_k]$.
\end{definition}

The dimensions in a relation space allows us to identify relations as follows: Suppose a set of attributes $A$ participates a query, then the dimensions it referenced is $K = A \cap D$. By our definition above, then there is at most one relation $\calF_j$ that $\calF_j \cap D = K$. Note that since $k$ can be less than $2^{|D|}$, this identification may find no relation, in which case the result is a {\sf NULL}. For example, suppose we have a relation space with two relations one with schema [A, B], and the other with schema [A, C], and dimensions are [A, B, C]. Then [A] identifies no relation. An important special case is RelationCube, which is a maximal relation space.
\begin{definition}[\textbf{RelationCube}]
\label{def:relation-cube}
A RelationCube object is a relation space where we have $k = 2^{|D|}$ relations, and every relation has all value attributes.
\end{definition}

Relation spaces serve two roles in MRA: They are usually the starting point of doing analysis. Or they can be obtained from MRA operator results.

We can create a relation space using the {\sf CreateRelationSpace} operator as follows:
\begin{lstlisting}[language=python,mathescape=true]
$\Psi$ = CreateRelationSpace(
  base_table,
  grouping_sets=cube([Device, Browser, Date]),
  aggreations={Cost:SUM, Cpc:SUM(Cost)/SUM(Clicks)},
  materialization=False,
)
\end{lstlisting}

Here {\sf CreateRelationSpace} is the analog of {\sf GROUP BY GROUPING SETS}, which performs aggregation w.r.t. multiple different grouping sets, but stores the results in a relation space, instead of a single (padded) relation. As we can see, relation spaces have the following advantages:
\begin{itemize}
    \item Compared with {\sf GROUP BY CUBE}, a relation space retains the schema information, and in fact one can use dimensions to identify different relations in the space. Perhaps more importantly, this allows us to control materialization exactly at the table granularity, so we will not be forced to materialize all of the data.
    \item Compared with creating all tables individually, one have a single logical entity -- namely the relation space -- to manage all the relations within the space. This management does not require us to assign names to relations, but can be achieved using combinations of dimensions, which is more natural for the analytical tasks described.
\end{itemize}


{\bf SliceRelation}. We now turn to problems related to data analysis, where the goal is to evaluate features for each entity in a structured and composable way. To support this, MRA introduces a second data type called {\sf SliceRelation}. While {\sf RelationSpace} helps organize multiple relations at varying granularities, {\sf SliceRelation} restructures the data around entities to support per-entity analysis. This design is conceptually related to the nested relational model, as it allows relation-valued attributes, but MRA constrains nesting to a single level, making the abstraction more tractable and well-suited for analytical tasks. To motivate this, suppose we begin with a {\sf RelationSpace} containing a [Device, Date, Cost, Cpc] relation and a [Device, Cost, Cpc] relation. We then: {\bf (1)} partition the [Device, Date, Cost, Cpc] relation by Device, without aggregation, producing one group per device, where each group contains a [Date, Cost, Cpc] relation; {\bf (2)} partition the [Device, Cost, Cpc] relation similarly to get per-device [Cost, Cpc] relations; and {\bf (3)} align these partitions on the Device attribute, producing tuples of the form shown in Table~\ref{tab:slice-vis}, where each tuple is keyed by a region (e.g., (Device=Pixel)) and contains a fixed set of relation-valued columns, each representing a feature relation. We call these structured tuples {\em slice tuples}, and the full collection forms a {\sf SliceRelation}. This representation enables analysts to express complex analytical logic, such as anomaly detection or feature filtering, over multi-granular data in a modular and declarative manner.

\begin{table}[htb]
\centering
\vspace{-0.1in}
\begin{tabular}{ccc}
    \Bigg(\parbox[c]{0.13\linewidth}{\centering (Device=Pixel)}, & 
    \begin{minipage}{0.25\linewidth}
        \centering
        \begin{tabular}{|l|l|l|}
            \toprule
            Date & Cost & Cpc \\
            \hline
            \hline
            2025-01-01 & 100 & 20 \\
            \hline
            2025-01-02 & 200 & 5 \\
            \hline
            ... & ... & ... \\
            \bottomrule
        \end{tabular},
    \end{minipage}
    &
    \begin{minipage}{0.15\linewidth}
        \centering
        \begin{tabular}{|c|c|}
            \toprule
            Cost & Cpc\\
            \hline
            \hline
            1500 & 20\\
            \bottomrule
        \end{tabular}
    \end{minipage}\Bigg)
\end{tabular}
\caption{\textbf{A slice tuple from arranging rows in the groupby cube results with respect to device Pixel. We call the tuple (Device=Pixel) a region; and there are two feature tables: one has schema [Date, Cost, Cpc], computed by the GROUP BY CUBE query with grouping [Device, Date], and the other has schema [Cost, Cpc], computed by the GROUP BY CUBE query with grouping [Device]. In particular, these rows correspond to different aggregation levels, but they are arranged into the same slice tuple.}}
\label{tab:slice-vis}
\end{table}

A set of slice tuples is then called a \emph{slice relation} (similar to that we call a set of tuples a relation in the classic relational model). This data representation can be computed using the {\sf Represent} operator in the multi-relational algebra (Section~\ref{sec:represent}):

\begin{lstlisting}[language=python,mathescape=true]
slice_relation = Represent(
  $\Psi$,
  region_schemas=[[Device],],
  feature_schemas=[[Date, Cost, Cpc], [Cost, Cpc],],
)
\end{lstlisting}
The result of {\sf Represent} can be visualized as a tabular structure as follows:
\begin{table}[htb]
  \centering
  \begin{tabular}{|l|c|c|}
    \toprule
    Region & [Date, Cost, Cpc] & [Cost, Cpc] \\\hline\hline
    (Device=Pixel) & $X^1_1$ & $X^1_2$ \\ \hline
    (Device=iPhone) & $X^2_1$ & $X^2_2$ \\ \hline
    ... & ... & ... \\\hline
    \bottomrule
  \end{tabular}
  \caption{\textbf{The tabular form of a slice relation computed by the Represent operator above.}}
  \label{tab:represent-visualization}
\end{table}

Note the following for a slice relation:
\begin{itemize}
\item Given a relation-space, a slice tuple may draw data \emph{from different relations} in the relation-space. For example, the data for daily Cpc timeseries (the first feature schema) comes from the relation with the schema [Device, Date, Cpc, Cost]. The Cost data (the second feature schema) comes from the relation with the schema [Device, Cpc, Cost].

\item In the {\sf Represent} operation, we identify proper relations using \emph{dimensions}: the region space [Device] and the first feature schema [Date, Cpc] gives a schema [Device, Date, Cpc], whose dimensions are [Device, Date], therefore we pick the relation with schema [Device, Date, Cpc, Cost]; the region space [Device] and the second feature schema [Cost] gives a schema [Device, Cost], whose dimensions are [Device], therefore we pick the relation with schema [Device, Cpc, Cost].

\item The analysis usually does not need all the data from a slice relation. For example for \textbf{(T1)}, the timeseries analysis does not need Cost column, and the cost check does not need the Cpc column.

\item A slice relation can be viewed as clustering the data points with respect to regions, but do not perform any further computation, such as aggregation, over the clustered data. A slice relation can thus be thought of as a representation of part of a relation-space to fit downstream analyses. Note that the clustered form can be analyzed in different ways in follow-up analyses, such as applying different timeseries analyses in this case. Therefore a clustered representation as such has potential performance benefits.
\end{itemize}

With slice relations, solving \textbf{(T1) -- (T4)}, and in fact all those variants in the \emph{varying entity granularity} case, become a \emph{single scan of a slice relation and filter}. For example for \textbf{(T1)}, we can first project to region schema [Device], followed by a {\sf SliceTransform} operation, and finally a {\sf SliceSelect} operation. For example for \textbf{(T1)}:
\begin{lstlisting}[language=python,mathescape=true]
S0 = SliceProject(
  slice_relation,
  region_schemas=[[Device],],
  feature_schemas=*,
)

S1 = SliceTransform(
  S0,
  slice_transformations=[
    # For the [Date, Cost, Cpc] feature column, we apply anomaly detection.
    ([Date, Cost, Cpc], CausalImpactAnalysis(time=Date, metric=Cpc)),
    # For the [Cost, Cpc] feature column, we extract the (total) cost feature.
    ([Cost, Cpc], Feature(attribute=Cost, alias=TotalCost)),
  ],
)

S2 = SliceSelect(
  S1,
  predicates=[TotalCost>1000],
)
\end{lstlisting}
where the {\sf CausalImpactAnalysis} computes a [Date, Cpc, IsAnomaly] table, and {\sf Feature(attribute=Cost, alias= TotalCost)} computes a [TotalCost] table. {\sf S2} is a slice relation, and if we want to flatten it into a relation space, we can then do

\begin{lstlisting}[language=python,mathescape=true]
$\Psi_1$ = Flatten(S2)
\end{lstlisting}

which computes a relation space with two relations in it: One with schema [Device, Date, Cpc, IsAnomaly], the other with schema [Device, TotalCost].

If one only cares about the end-to-end transformation from a relation space to another, we can also compose the above MRA operators, {\sf Represent}, {\sf SliceTransform}, {\sf SliceSelect}, and {\sf Flatten}, as shown in the following sequence:
\begin{lstlisting}[language=python,mathescape=true]
S0 = Represent(
  $\Psi$,
  region_schemas=[[Device],],
  feature_schemas=[[Date, Cpc], [Cost],],
)

S1 = SliceTransform(
  S0,
  slice_transformations=[
    CausalImpactAnalysis(time=Date, metric=Cpc),
    Feature(attribute=Cost, alias=TotalCost),
  ],     
)

S2 = SliceSelect(
  S1,
  predicates=[TotalCost>1000],
)

$\Psi_1$ = Flatten(S2)
\end{lstlisting}

We can thus form a mega-operator called {\sf Crawl}, which directly operates on a relation-space and outputs another relation space,
which allows us to write a single query for each task:
\begin{lstlisting}[language=python,mathescape=true,label={lst:crawl-1}]
$\Psi_1$ = Crawl(
  $\Psi$,
  region_schemas=[[Device],],
  slice_transformations=[
    CausalImpactAnalysis(time=Date, metric=Cpc),
    Feature(attribute=Cost, alias=TotalCost),
  ],
  predicates=[TotalCost>1000,],
)
\end{lstlisting}

\subsection{More follow-up analysis leveraging MRA's composability}
\label{sec:more-composability}

In the previous section we described how to use MRA to solve a set of entity-search analytics tasks, and compared with SQL. In practice, there are more follow-up analysis an analyst might want to perform on the output of crawl. The output of a crawl is a relation space, therefore by the composability of MRA, we can again apply MRA to the results. There are however two broad categories of analysis one might need to do:
\begin{itemize}
\item \textbf{Discarding the schema structure of a relation space}. There are natural cases where we want to {\sf UNION ALL} relations in a relation space, get a single relation, view it as a base relation, and then create a new relation space from aggregating the new base relation, and then apply more MRA operators to transform the new relation space. This gives even higher-order analysis, similar to ``aggregation over aggregation'' situation in SQL.

\item \textbf{Respect the schema structure of a relation space}. There are also natural cases where we want to keep the schema structure of the relation space, and do more analysis respecting the schema structure.
\end{itemize}

In the rest of the section, we give one example for each category.

\textbf{Follow-up Analysis 1.} Suppose we are interested in analyzing those dates where we have an anomaly. We can do this with a traditional relational algebra Select operator with the predicate {\sf IsAnomaly=True}. Let us call the resulting relation space $\Psi_2$. Now in $\Psi_2$, we may found that Device=Pixel appears frequently in different entities with different granularity. For example, we might see it appear in the [Device] entity group, with (Device=Pixel), and we might also see it appear in multiple device-browser entities, say (Device=Pixel, Browser=Chrome), (Device=Pixel, Browser=Edge). In other wise, Pixel is a \emph{frequent pattern} found in the previous crawl analysis. Users might want to identify such frequent patterns by examining $\Psi$. This can be achieved again using MRA.

In order to do so, we first flatten $\Psi_1$ into a single relation, by {\sf UNION ALL} all relations in the space. Let the result be $X_2$. This is an important step because we want to analyze \emph{across} different entity granularity in $\Psi_2$. Then, we create another relation space $\Psi_3$, by aggregating the counts over [Device], [Browser], [Device, Browser]:
\begin{lstlisting}[language=python,mathescape=true]
$\Psi_3$ = CreateRelationSpace(
  $X_2$,
  grouping_sets=cube([Device, Browser]),
  aggreations={Count:COUNT(1)},
  materialization=False,
)
\end{lstlisting}

For example, for the grouping set = [], the {\sf Count} is the total count of all anomalies. For grouping set=[Device], the row with (Device=Pixel) records the count of anomalies where Device=Pixel appears (e.g. (Device=Pixel) is anomalous in $\Psi_1$, (Device=Pixel, Browser=Chrome) is anomalous in $\Psi_1$, and so on). Now we can find frequent patterns using the classic frequent itemset mining by computing the {\sf support}.

\begin{lstlisting}[language=python,mathescape=true,label={lst:crawl-2}]
$\Psi_4$ = Crawl(
  $\Psi_3$,
  region_schemas=[[Device], [Browser], [Device, Browser],],
  slice_transformations=[
    FrequentItemsetAnalysis(count_column='Count'),
  ],
  predicates=[support>.4,],
)
\end{lstlisting}

\begin{figure}[htb]
\centering
\includegraphics[width=.3\linewidth]{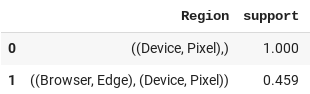}
\caption{{\bf Results of executing pipe~\ref{lst:crawl-2}}. We found the Device=Pixel to be a frequent pattern with support 1
}%
\label{fig:pipe2-results}
\end{figure}

The results are shown in Figure~\ref{fig:pipe2-results}. For example, (Device=Pixel) is found as the most frequent pattern with support=1.0

\textbf{Follow-up Analysis 2.} In the previous follow-up analysis, we need to discard the schema structure of $\Psi_1$, since we want to aggregate across different entity granularity. There are also cases where we need to analyze in a way that \emph{respects} the schema structure. In this section we give an example. Suppose that for the date-cpc time series identified in $\Psi_1$, we want to understand the correlation with budget for each of the entities. The budget timeseries for different entities are stored in another relation space $\widetilde{\Psi}$, with relations with schema [Device, Date, Budget], [Browser, Date, Budget], [Device, Browser, Date, Budget]. The correlations can be analyzed as follows.

We first represent $\Psi_1$ and $\widetilde{\Psi}$ into entities with timeseries features:
\begin{lstlisting}[language=python,mathescape=true]
S1 = Represent(
  $\Psi_1$,
  region_schemas=[[Device], [Browser], [Device, Browser],],
  feature_schemas=[[Date, Cpc],],
)

S2 = Represent(
  $\widetilde{\Psi}$,
  region_schemas=[[Device], [Browser], [Device, Browser],],
  feature_schemas=[[Date, Budget],],
)
\end{lstlisting}

Then we can analyze the correlations at the slice level, using SliceJoin.
\begin{lstlisting}[language=python,]
S3 = SliceJoin(
  S1,
  S2,
  join_conditions=REGION_EQUAL,
)
\end{lstlisting}

This gives a slice relation with region schemas [Device], [Browser], [Device, Browser], and feature schemas [Date, Cpc] and [Date, Budget].
We can then use {\sf SliceInternalJoin} to join the two feature tables:

\begin{lstlisting}[language=python,]
S4 = SliceInternalJoin(
  S3,
  feature_schemas=[[Date, Cpc], [Date, Budget]],
  join_conditions=[[Date, Cpc].Date=[Date, Budget].Date],
)
\end{lstlisting}

Finally, we can crawl for correlations:
\begin{lstlisting}[language=python,mathescape=true]
Crawl(
  S4,
  region_schemas=[[Device], [Browser], [Device, Browser],],
  slice_transformations=[
    CorrelationAnalysis(
      metric1=Cpc,
      metric2=Budget,
      entity_specifiers=[Date],
    ),
  ],
  predicates=[support>.4,],
)
\end{lstlisting}


\section{Data Model}
\label{sec:data-model}

\subsection{RelationSpace}


Given a relation-space $\Psi$ and a relation $X$ in $\Psi$, we use $\dim_\Psi(X)$ to denote the dimensions of $X$ w.r.t. $\Psi$. That is $\dim_\Psi(X) = \schema(X) \cap D$. When $\Psi$ is clear from the context, we simply denote $\dim(X)$. A key property of a RelationSpace is that dimension set can be used to identify relations: if a set of attributes $A$ participates a query, then we can use $A \cap D$ to uniquely identify at most one table with schema $\calF$ such that $\calF \cap D = A \cap D$ (i.e., their dimensions match). Note that a RelationSpace can have at most $2^{|D|}$ relations.

Note that compared to OLAP cubes, RelationSpace is much less constrained: for example, we don't require any aggregation relationship among the relations stored in a relation-space, in particular, they don't have to be aggregated from one underlying ``base table''.

\begin{example}[\textbf{A single relation}]
    A single relation can be casted into a relation space with empty dimensions (i.e., {\sf dimensions=[]}). That is, we don't need any dimension to identify this relation.
\end{example}

We can also easily turn a relation space back to a single relation

\begin{example}[\textbf{Converting a relation space to a single relation}]
    Given a relation space $\Psi$, we can convert it into a single relation by {\sf UNION ALL} all relations in it. Note that in view of standard SQL, this will have {\sf NULL} padding.
\end{example}

\subsection{SliceRelation}

In the rest of this section, we formalize the data type SliceRelation. To do so, we first formalize RawSliceRelation, which are slice relations without dimensions.

\subsubsection{RawSliceRelation: Slice relations without dimensions}
We first define RawSliceRelation, which is a data type for slice relations without dimensions. We begin with a function definition of relations in the classic relational algebra. The standard definition is that a relation is a set of tuples. However for our purpose of generalization, it is better to define it by a function. Specifically, let $X$ be a relation with $n$ rows. Then we can define a function $f_X$ with domain $[n] \times \mathcal{A}$, where $[n] = \{1, 2, \dots, n\}$ and $\mathcal{A}$ is the set of attributes of $X$, and co-domain $\mathcal{V}$, which are values that attributes can take.
Here $[n]$ models the primary key of a table, that is row index $1,2,3,\dots$. Then, given an integer $i \in [n]$, $a \in \mathcal{A}$, $f_X(i, a)$ gives the value of attribute $a$ in the row $i$. With this definition, the $i$-th tuple in a relation is $f_X(i, \cdot)$, where we fix the first parameter as $i$, and leave the second parameter free. In particular, a tuple is also a function, of type $\mathcal{A} \rightarrow \mathcal{V}$. For example, suppose the third row of $X$ is a tuple (Device=Pixel, Browser=Chrome), then the tuple is the function $g(\cdot)=f_X(3, \cdot)$, where $g$(Device)=Pixel, $g$(Browser)=Chrome.

We use Relation to denote the data type of relations. We now define SliceRelation, a data type that generalizes Relation. We set up some terminologies and notations first.

\textbullet\: Let $U$ be a universe of attributes. A region schema is a subset of $U$. A feature schema is also a subset of $U$. 

\textbullet\:  We use $\overline{\Gamma}=[\Gamma_1$, \dots, $\Gamma_m]$ to denote a set of $m$ region spaces, and use $\calF_1, \dots, \calF_k$ to denote $k$ feature schemas. We use $\overline{\calF}$ to denote the set of schemas $[\calF_1, \dots, \calF_k]$ (in particular, since $\overline{\calF}$ is a set, all the feature schemas are distinct).

\textbullet\:  With these, the attributes of different regions schemas are $\calA_r = \cup_{i \in [m]}\Gamma_i$ and the attributes of different feature schemas are $\calA_f = \cup_{i \in [k]}\calF_i$.

\textbullet\:  We say that $\langle \overline{\Gamma}, \overline{\calF} \rangle$ form a legitimate raw slice-relation schema if $\calA_r \cap \calA_f = \emptyset$.

\textbullet\:  We use the notation $\Omega[\overline{\Gamma}]$ to denote the set of tuples whose schema is in $\overline{\Gamma}$. For example, suppose $\overline{\Gamma}=$[[Device], [Browser]], then $\Omega[\overline{\Gamma}]$ contains tuples such as (Device=Pixel) and (Browser=Chrome).

\textbullet\:  We use the notation $\calT[\calF]$ to denote the set of relations whose schema is $\calF$. For example,  $\calT[[{\rm Date}, {\rm Cost}]]$ denotes the set of relations whose schema is [Date, Cost]. Similar to our notations for regions spaces, we use $\calT[\overline{\calF}]$ to denote the set of relations whose schema is an element of $\overline{\calF}$.  

\begin{definition}[\textbf{RawSliceRelation}]
\label{def:slice-repr-multi}
Let $U$ be a universe of attributes, $\overline{\Gamma} = [\Gamma_1, \dots, \Gamma_m]$ be a set of $m$ region spaces, and $\calF_1, \dots, \calF_k$ be $k$ feature schemas. Denote $\overline{\calF} = [\calF_1, \dots, \calF_k]$. Suppose $\overline{\Gamma}$ and $\overline{\calF}$ form a legitimate slice-relation schema. A RawSliceRelation $\Lambda$ is a function
$$f_\Lambda: \calR \times \overline{\calF} \rightarrow \calT[\overline{\calF}],$$
where $\calR$ is a finite subset of $ \Omega[\overline{\Gamma}]$, and for each $r \in \calR, \calF \in \overline{\calF}$, $\schema(f_\Lambda(r, \calF)) = \calF$.

\end{definition}

That is, compared to a usual function definition of relation:

\textbullet\: We replace the first component of the function, the integer primary key $i \in [n]$, by a tuple $r \in \Omega[\overline{\Gamma}]$ (which we call the region), and

\textbullet\:  We replace the second component of the function, an attribute $a \in {\cal A}$, by a set of attributes $\calF$ (which we call the feature schema)

\begin{example}
Let $U$=[Device, Browser, Date, Cost], $\overline{\Gamma} = [[{\rm Device}], [{\rm Browser}],]$, and $\overline{\calF} = [[{\rm Date}, {\rm Cost}],]$ (there is only one feature schema).

\textbf{(1)} Let $\calR$ be a finite set of tuples whose schemas are from $\overline{\Gamma}$. Then we may have tuple (Device=Pixel) in $\calR$, also tuples such as (Browser=Chrome).

\textbf{(2)} Let $\calT$ be the set of relations whose schema are from $\overline{\calF}$. In this case, since we only have one feature schema, all the relations in $\calT$ have schema [Date, Cost].

\textbf{(3)} Then a  raw slice relation is of type $\calR \times \overline{\calF} \rightarrow \calT$. That is, we take a region $r \in \calR$, and a schema $\calF \in \overline{\calF}$, and the slice function maps these two to a table whose schema is $\calF$. Suppose $r=$(Device=Pixel), and there is only one schema in $\overline{\calF}$, that is $\calF$=[Date, Cost], so we map (Device=Pixel) and [Date, Cost], that is $f_\Lambda(\text{(Device=Pixel)}, \text{[Date, Cost]})$,  to a relation of schema [Date, Cost].
\end{example}


A \emph{slice tuple} of a slice relation function is thus a function where we fix the first parameter as some region, that is, $f_\Lambda(r, \cdot)$, where $r$ is a region. Note that regions can have different schemas, and we call different schemas for regions \emph{region schemas}. A slice tuple is thus of the type $\overline{\calF} \rightarrow \calT$. 

%

We can also define raw slice relation using a pure set theoretic language. First, with the above notations, we can form the following cross product:
$$\calK = \Omega[\overline{\Gamma}] \times \calT[\calF_1] \times \calT[\calF_2] \times \cdots \times \calT[\calF_k]$$
Then a slice relation can be defined as follows:
\begin{definition}
Let $\langle\overline{\Gamma}, \overline{\calF}\rangle$ be a legitimate raw slice-relation schema. Then a slice relation $\Lambda$ is a finite subset of $\calK$ where the first component is a primary key. In other words, there is a functional dependency $r \rightarrow (X_1, X_2, \dots, X_k)$ for $(r, X_1, \dots, X_k) \in \Lambda$.
\end{definition}

In the tabular form, a slice relation can be visualized as follows:
\begin{table}[htb]
  \centering
  \begin{tabular}{|l|l|l|l|l|}
    \toprule
    Region & $\calF_1$ & $\calF_2$ & $\cdots$ & $\calF_k$\\\hline\hline
    $r_1$ & $X^1_1$ & $X^1_2$ & $\cdots$ & $X^1_k$ \\ \hline
    $r_2$ & $X^2_1$ & $X^2_2$ & $\cdots$ & $X^2_k$ \\ \hline
    ... & ... & ... & ... & ... \\\hline
    \bottomrule
  \end{tabular}
  \caption{\textbf{The tabular form of a slice relation (i.e. an instance of data type SliceRelation)}. We have region spaces $\Gamma_1, \dots, \Gamma_n$, and feature schemas $\calF_1, \dots, \calF_k$. The `Region' column is a primary key.
  }
  \label{tab:slice_repr-multi-feature-sets}
\end{table}

\textbf{Terminologies and notations}. In the tabular form, each feature table column is indexed by a feature schema $\calF_i$; such a column differs from a traditional column, which is indexed by an attribute. Therefore, if needed, we will refer to such a column as a ``feature table column''. The ``Region'' column is the primary key, which takes value of a region tuple. A slice tuple can then be denoted as:
$\langle \region=r: \calF_1=X_1, \calF_2=X_2, \dots, \calF_k=X_k \rangle,$
where $r$ is a region, and $X_i$'s are feature tables, and the colon symbol is used to indicate that the region $r$ is a primary key (this is consistent with the use of the colon symbol in a dictionary). This notation is similar to our notation for traditional tuples, such as (Device=Pixel, Browser=Chrome), where `Device' and `Browser' are attributes, and `Pixel' and `Chrome' are values of the attributes. In the context where feature schemas are clear from the context, we use a shorthand $r: (X_1, X_2, \dots, X_k)$. Finally, we define a \emph{slice} to be a pair $(\text{Region}=r, \calF=X)$ (or shorthand $(r, X)$).

\begin{definition}[\textbf{Slice Schema Block}]
\label{def:block}
A slice schema block, or schema block, or just block, is a pair of schemas $(\Gamma, \calF)$. $\Gamma$ is the schema of regions, and $\calF$ is the schema of feature tables
\end{definition}
A slice relation corresponding to a single slice schema block is called a \emph{slice relation block}. Table~\ref{tab:slice-relation-block} visualizes a slice relation for a single block.
\begin{table}[htb]
  \centering
  \vspace{-0.1in}
  \begin{tabular}{|l|l|l|l|l|}
    \toprule
    Region & $\calF$\\\hline\hline
    $r_1$ & $X_1$ \\ \hline
    $r_2$ & $X_2$ \\ \hline
    ... & ... \\\hline
    \bottomrule
  \end{tabular}
  \caption{\textbf{Slice relation of a single schema block}.
  $\Gamma$ is the region space, and $\calF$ is the feature schema.
  }
  \label{tab:slice-relation-block}
\end{table}

\subsubsection{Represent: Representing a relation-space as a (raw) slice-relation}
\label{sec:represent}
Represent operator represent a part of a relation-space into a slice relation with non-trivial region schemas, and thus forming appropriate clustering of data, as we motivated earlier in Table~\ref{tab:slice-vis}. 
Concretely, recall that we consider the query ``find devices that have timeseries Cpc anomalies and total revenue > 1000''. With a relation-space with schemas [Cpc, Revenue], [Device, Cpc, Revenue], [Date, Cpc, Revenue], and [Device, Date, Cpc, Revenue], we would like to group the [Device, Date, Cpc, Revenue] table by Device and consider [Date, Cpc] as feature table, and group the [Device, Cpc, Revenue] table also by Device and consider the [Revenue] table. This is captured by a single {\sf Represent} operation as:

\begin{lstlisting}[language=python,mathescape=true]
Represent(
  $\Psi$,
  region_schemas=[[Device],],
  feature_schemas=[[Date, Cpc], [Revenue],],
)
\end{lstlisting}
This operation declares that we want to make Device tuples as regions, and for each region we consider two feature tables [Date, Cpc] and [Revenue]. If we consider the region space and the first feature schema, we get a schema [Device, Date, Cpc], and so it is clear that we should match the table with schema [Device, Date, Revenue, Cpc] in the relation-space, and make pull Device into regions. What about the second feature schema? We get [Device, Revenue] and there is now an issue that both tables in the relation-space have [Device, Revenue] as columns, so which table we should match to? (note that we want to match the second). The answer should be simple: We use dimensions! That is, In our previous example, [Device, Date] is a dimension set, This is because $D \cap \calF_1 = \text{[Device, Date]}$, $D \cap \calF_2=\text{[Device]}$. Therefore, [Device] uniquely identifies the table with schema [Device, Revenue], and [Device, Date] identifies uniquely the table with schema [Device, Date, Revenue] (note however [Date] and [] identify no table in the relation-space). By contrast, $D=\text{[Device]}$ is not a dimension set, this is because $D \cap \calF_1 = \text{[Device]} = D \cap \calF_2$.

\begin{definition}[\textbf{Represent a Slice Relation Block}]
\label{def:slice-repr}
Let $\Psi$ be a relation-space with dimension set $D$, and $(\Gamma, \calF)$ be a slice block. The {\sf Represent} is defined as
\begin{lstlisting}[language=python,mathescape=true]
Represent(
  $\Psi$,
  region_schema=$\Gamma$,
  feature_schema=$\calF$, 
)
\end{lstlisting}

and is computed as follows:

\textbf{(1)} \textbf{Table matching}. Let $\calT = \Gamma \cup \calF$, and $K = \calT \cap D$. Since $D$ is a dimension set, $K$ uniquely identifies at most one schema. If there is no table identified, we return {\sf NULL}. Otherwise, let the identified table be $Z$ in $\Psi$ with $\schema(Z) = \calF_i$.  We then project $Z$ to $\calT$, i.e., $Z|_{\calT}$.

\textbf{(2)} \textbf{Partition to get slices}. Finally, we partition $Z|_{\calT}$ over $\Gamma$, and get a set of slices $(r, X)$, where $\schema(r) = \Gamma$ and $\schema(X) = \calF$. Table~\ref{tab:block-represent} gives a visualization of the result of represent.
\end{definition}
\begin{table}[htb]
  \centering
  \vspace{-0.1in}
  \begin{tabular}{|l|l|l|l|l|}
    \toprule
    Region & $\calF$\\\hline\hline
    $r_1$ & $X_1$ \\ \hline
    $r_2$ & $X_2$ \\ \hline
    ... & ... \\\hline
    \bottomrule
  \end{tabular}
  \caption{\textbf{Representing a block}. $\Gamma$ is the region space and $\calF$ is the feature set.
  }
  \label{tab:block-represent}
\end{table}

\begin{figure}[htb]
  \centering
  \begin{subfigure}[c]{.28\textwidth}  
  \centering
  \includegraphics[width=\linewidth]{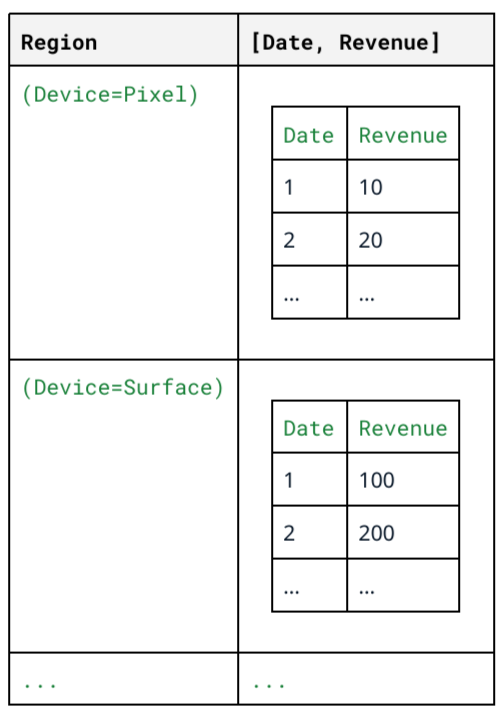}
  \caption{
    \scriptsize $\Gamma$=[Device], $\calF$=[Date, Revenue].
  }
  \label{fig:represent-ex1}
  \end{subfigure}\quad
  \begin{subfigure}[c]{.28\textwidth}  
  \centering
  \includegraphics[width=\linewidth]{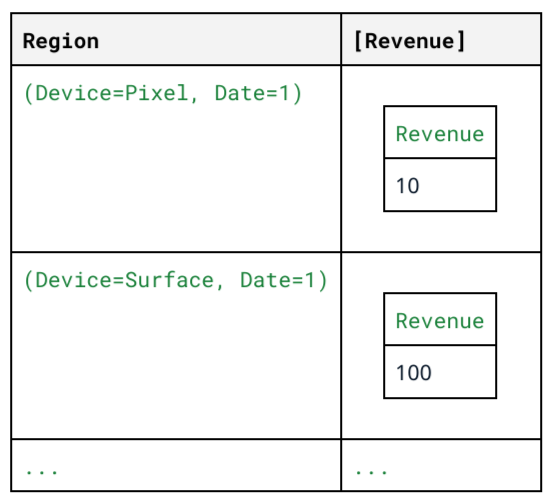}
  \caption{
    \scriptsize $\Gamma$=[Device, Date], $\calF$=[Revenue].
  }
  \label{fig:represent-ex2}
  \end{subfigure}
  \begin{subfigure}[c]{.28\textwidth}  
  \centering
  \includegraphics[width=\linewidth]{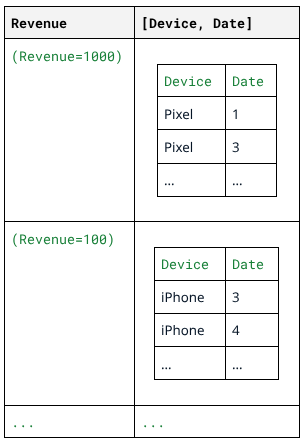}
  \caption{
    \scriptsize $\Gamma$=[Revenue], $\calF$=[Device, Date].
  }
  \label{fig:represent-ex3}
  \end{subfigure}
  \caption{Different representations of the same table with schema [Device, Date, Revenue]. Figure~(\ref{fig:represent-ex3}) shows that one doesn't need to have any dimension attributes in regions.}
\end{figure}

Importantly, we can represent the same table in a multi-relation as different slices, the following gives an example.
\begin{example}[\textbf{Representing the same table as different slices}]
First,
$$\represent(\Psi, \text{region\_schema=[Device]}, \text{feature\_schema=[Date, Revenue]})$$
matches the table [Device, Date, Revenue], and computes a set of slices where region has schema [Device], and the table has schema [Date, Revenue]. The result is shown in Figure~\ref{fig:represent-ex1}.
By contrast,
$$\represent(\Psi, \text{region\_schema=[Device, Date]}, \text{feature\_schema=[Revenue]})$$
also macthes the table [Device, Date, Revenue], but computes a set of slices where region has schema [Device, Date], and the features have schema [Revenue]. The result is shown in Figure~\ref{fig:represent-ex2}.
\end{example}

With the block representation, Represent is then defined as follows for multiple region spaces and feature schemas:
\begin{definition}[\textbf{Representing with multiple region spaces and feature schemas}]
Let $\overline{\Gamma}$ be a set of region spaces and $\overline{\calF}$ be a set of feature schemas. $\represent$ receives $\overline{\Gamma}$, $\overline{\calF}$, and a dimension set $D$, and computes as follows: For each $\Gamma_i$ and $\calF_j$ we do a block representation, then we join different slices from different block representation into one with equi-join on the region, (i.e., for $r: X_1$, $r: X_2$, $\dots$, $r: X_k$,  we merge them into one slice tuple $r: (X_1, \dots, X_k)$). The resulting set of slice tuples is the output slice relation. 
\end{definition}

\textbf{Succinct specification}. Asking users to manually specify every region space can be cumbersome. There are various ways to make the specifications more succinct. For example, we can use {\sf cube([Device, Browser])} to encode all non-trivial subsets of [Device, Browsers] for region schemas, that is [Device], [Browser], [Device, Browser]. Therefore cube with $n$ attributes encodes $2^n-1$ region spaces.

\subsubsection{Flattening with dimensions}
\label{sec:flatten}
$\represent$ structures a multi-relation to a region-indexed slice-relation. What about the reverse? To this end, we observe an immediate issue: Suppose we do the following $\represent$:
\begin{lstlisting}[language=python,mathescape=true]
Represent(
  $\Psi$,
  region_schemas=[[Device],],
  feature_schemas=[[Date, Cpc], [Date, Revenue],],
)
\end{lstlisting}

Then the column Date is repeated in two different feature tables (i.e., the data is de-normalized). Further, since we do projections during $\represent$, these two feature tables may not have the same number of dates (for example, some dates may be missing in the [Date, Revenue] table if there are NULL values). Therefore, we need to know that Date was a dimension in the multi-relation before representation, and also, when we join [Date, Cpc] and [Date, Revenue] on Date, we should do outer-join. This motivates that we keep track of the dimension information within a slice-relation, which is not book-kept in the RawSliceRelation. Formally, we observe that with dimensions, flattening is simple due to an equivalence relation induced by a set of dimensions.

\begin{definition}[\textbf{Equivalence relation $\sim_D$ induced by a dimension set $D$}]
Let $D$ be a dimension set. Then for a slice tuple $r : (X_1, . . . , X_k)$, we say that $X_i \sim_D X_j$ if $\dim(X_i) = \dim(X_j)$. If $D$ is clear from the context, we simply write $\sim$, instead of $\sim_D$.
\end{definition}
$\sim$ is an equivalence relation (reflexive, symmetric, and transitive). Now, the Flatten operation is defined as follows.

\begin{definition}[\textbf{Flatten}]
Let $\langle \overline{\Gamma}; \calF_1, \dots, \calF_k\rangle; D$ be a slice relation schema with region schemas $\overline{\Gamma}$, feature schemas $\calF_1, \dots, \calF_k$, and dimension set $D$. The $\flatten$ operation receives a slice relation $\Lambda$ with this schema, and compute as follows:

\textbf{(1) Computing new slice tuples using dimensions}. For each slice tuple $r: (X_1, \dots, X_k)$, we partition $X_1, \dots, X_k$ into equivalence classes under $\sim$, then within each equivalence class, we perform a full-outer join; performing these joins gives a set of relations $(Z_1, \dots, Z_l)$. Doing this transformation for all slice tuples gives us a new slice relation $\Psi'$, which has schema $\langle \overline{\Gamma}, \overline{\calF}' \rangle$, where $\overline{\calF}' = [\calF'_1, \dots, \calF'_{l}]$.

\textbf{(2) Moving regions into features}. For each $\Gamma_i$ and $\calF'_j$, we create a relation of schema $\calT_{ij} = \Gamma_i \cup \calF'_j$. The resulting collection of tables $\{T_{ij}\}$ forms a multi-relation.

\textbf{(3) Dimensions for the output multi-relation}. Finally, the dimension set for the output multi-relation is $D$.
\end{definition}

As a concrete example, consider a slice tuple $(r=(), Z_1, Z_2, Z_3)$, where schema($Z_1$)=[Date, Cpc], schema($Z_2$)=[Date, Revenue], and schema($Z_3$)=[Cpc]. We use $D=\text{[Device, Date]}$, then we have two equivalence classes, $[Z_1, Z_2]$ since $\dim(Z_1)=\dim(Z_2) = \text{[Date]}$ and $[Z_3]$ (as $\dim(Z_3) = []$). Then we will join the tables in the first equivalence class on Date and get $Z_1'$, which has schema [Date, Cpc, Revenue]. The output slice tuple is $(r=(), Z_1', Z_2'=Z_3)$, which is already a multi-relation by throwing away the empty region tuple.

\subsubsection{SliceRelation: Slice relations with dimensions}

Astute readers will now realize a potential issues: Even though we say $D$ is a dimension set in the flattening process, there is really no guarantee that after flattening $D$ forms a dimension set. This is addressed in the definition: We force that $D$ must be a legitimate dimension set in the flattened table-space.
\begin{definition}[\textbf{SliceRelation}]
    A SliceRelation data type has schema $\langle D, V, \overline{\Gamma}, \overline{\calF}\rangle$, where $D$ is a set of dimensions. An instance of SliceRelation $\Lambda$ satisfy that $\flatten(\Lambda)$ is a legitimate relation-space with dimensions $D$.
\end{definition}

\section{Operations on slice relations}
\label{sec:operations-sr}
Now we define algebraic operations on slice relations. As we can see now, slice relations have a nested structure, where each feature table column contain nested tables. Therefore, there are two categories of operations:
\begin{itemize}
\item \textbf{Data Representation operations}. Similar to $\represent$, but which only converts a multi-relation to a region-indexed slice relation, we want a general representation operation that represent a slice relation to another slice relation.

\item \textbf{Internal operations}. We operate on the multiple feature tables nested within each slice tuple. Each slice tuple is transformed, but the existence of that slice tuple always remains. This gives operations of SliceInternalProject, SliceInternalSelect, and SliceInternalJoin. One can view these as the classic Project, Select, Join apply to feature tables within each slice tuple. Note however, SliceInternalProject and SliceInternalSelect apply to multiple tables simultaneously, generalizing Project and Select, which are applied to one table at a time.

\item \textbf{External operations}. We operate on the slice tuples from an external point of view, using the slice relation schemas. The slice tuples may be filtered due to failing some predicates. This gives operators SliceProject, SliceSelect, SliceJoin, and SliceTransform.
\end{itemize}

\subsection{Data representation operations}

\subsubsection{SliceRepresent}
\label{sec:slice-represent}

We have defined Represent operation from multi-relation to a non-trivial slice relation. We now generalize this to transform an arbitrary slice relation.

\begin{definition}[\textbf{SliceRepresent}]
{\sf SliceRepresent} operator is defined as
\begin{lstlisting}[language=python,mathescape=true]
SliceRepresent(
  $\Lambda$,
  region_schemas=[$\Gamma'_1, \Gamma'_2, \dots, \Gamma'_{m'}$],
  feature_schemas=[$\calF'_1, \calF'_2, \dots, \calF'_{k'}$],
)
\end{lstlisting}
is logically a composition of {\sf Flatten} followed by a {\sf Represent}:
\begin{lstlisting}[language=python,mathescape=true]
Represent(
  Flatten($\Lambda$),
  region_schemas=[$\Gamma'_1, \Gamma'_2, \dots, \Gamma'_{m'}$],
  feature_schemas=[$\calF'_1, \calF'_2, \dots, \calF'_{k'}$],    
)
\end{lstlisting}
\end{definition}

\begin{example}
For example, suppose we have a slice relation $\Lambda$ with region schemas, {\sf [Device]}, {\sf [Device, Date]}, and one feature schema {\sf [Revenue, Cpc]}. Then
\begin{lstlisting}[language=python,mathescape=true]
SliceRepresent(
  $\Lambda$,
  region_schemas=[[Device],],
  feature_schemas=[[Date, Cpc], [Revenue]], 
  dimensions=[Device, Date]],
)
\end{lstlisting}
will first flatten $\Lambda$ to a multi-relation with schemas {\sf [Device, Revenue, Cpc]} and {\sf [Device, Date, Revenue, Cpc]}, then represent it to a slice relation with region schema {\sf [Device]}, and two feature schemas {\sf [Date, Cpc]} and {\sf [Revenue]}.
\end{example}

\subsection{Internal operations}

\subsubsection{SliceInternalProject}
\label{sec:slice-internal-project}

SliceInternalProject projects each feature table column to a sub-schema.
\begin{definition}[\textbf{SliceInternalProject}]
\label{def:slice-internal-project}
The SliceInternalProject operator
\begin{lstlisting}[language=python,mathescape=true]
SliceInternalProject(
  $\Lambda$,
  projection_list=[
    $(\calF_1, \widehat{\calF}_1), (\calF_2, \widehat{\calF}_2), \dots, (\calF_l, \widehat{\calF}_l),$
  ],
)
\end{lstlisting}
where $\widehat{\calF}_i \subseteq \calF_i$ ($i \in [l]$), computes a slice relation by projecting feature table column $\calF_i$ to attributes $\widehat{\calF}_i$, such that:
\begin{itemize}
    \item a shorthand $\ast$ is provided, similar to SQL, so one can specify $(\calF, \ast)$ to keep the entire feature table column.
    \item if two projected feature schemas are the same, the projected relations are unioned.
    \item if a feature table schema is not specified in the projection list, that feature table column is ignored in the results.
\end{itemize}
\end{definition}

\subsubsection{SliceInternalSelect}
\label{sec:slice-internal-select}
SliceInternalSelect uses a select statement to filter feature tables within each slice tuple.
\begin{definition}[\textbf{SliceInternalSelect}]
\label{def:slice-internal-select}
The SliceInternalSelect operator
\begin{lstlisting}[language=python,mathescape=true]
SliceInternalSelect(
  $\Lambda$,
  select_statement,
)
\end{lstlisting}
where $\Lambda$ is a relation-space, {\sf select\_statement} is a SQL select statement which reference (only) feature schemas. It computes an output slice relation by applying the {\sf select\_statement} to the feature tables of each slice tuple in $\Lambda$. The select statement only filters feature tables, without changing any of the feature schemas. The output relation has the same slice relation schema with $\Lambda$, but with tuples filtered.
\end{definition}

\begin{figure}[htb]
\centering
\includegraphics[width=0.5\linewidth]{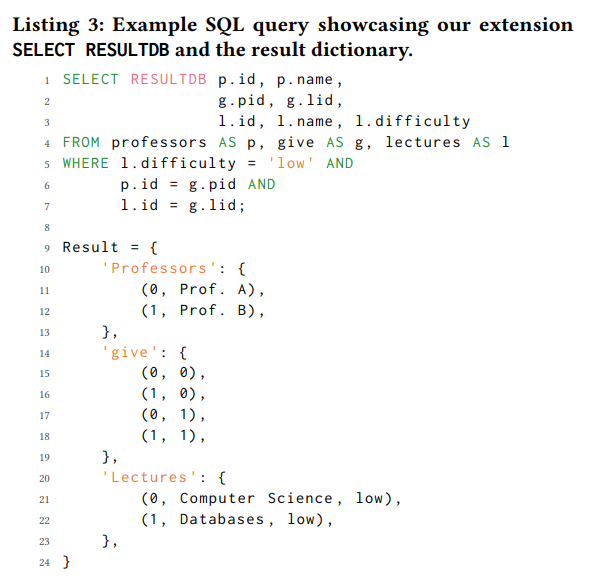}
\caption{Listing 3 from~\cite{nix2023sqlstatementreturneddatabase}. A select statement on multiple tables. This is a special case of SliceInternalSelect: It is a SliceInternalSelect applied to a MultiRelation.}
\label{fig:select-resultdb-listing3}
\end{figure}

\begin{example}[\textbf{{\sf SELECT RESULTDB} is a special case of SliceInternalSelect}]
\label{example:resultdb}
A recent work~\cite{nix2023sqlstatementreturneddatabase} advocates to extend SQL to compute multiple tables as output (of course, receiving multiple tables as input). Figure~\ref{fig:select-resultdb-listing3} gives an example of the proposed {\sf SELECT RESULTDB} query. We note the following:
\begin{itemize}
\item While the select clause looks like a normal SQL query, the first and foremost difference is that the output is different. Traditional relational algebra forces one to output a single table. Howevever, the {\sf SELECT RESULTDB} query returns {\sf Result} which consists of \emph{three} tables: one {\sf Professor} table, one {\sf give} table, one {\sf Lectures} table. For each table, we have rows of chosen columns. For example for the {\sf Professor} table, the {\sf SELECT RESULTDB} clause chooses the `p.id' and `p.name', that is we choose id and name columns from the original {\sf Professor} table, and the result has two rows, one with id=0, and name=`Prof. A', and the other with id=1, and name=`Prof. B'.

\item Note that in the WHERE clause we have two predicates, one is to choose l.difficulty='low', thus we only keep low difficulty lectures. The more interesting predicate is that we have `p.id=g.pid AND l.id=g.lid'. This will give semi-join conditions to filter rows that only appear in one table, but not others (for example, a professor that does not have any lecture to give)
\end{itemize}
We note that this query is a special case of MRA; specifically, this is a SliceInternalSelect applied to a RelationSpace, and the output is another RelationSpace that has no region.
\end{example}

\subsubsection{SliceInternalJoin}
\label{sec:slice-internal-join}

SliceInternalJoin joins tables in different feature table columns within a slice tuple. This is no different than SQL, except that now this is a batch operation: We join on the tables for each slice tuple.
\begin{definition}[\textbf{SliceInternalJoin}]
\label{def:slice-internal-join}
The SliceInternalJoin operator
\begin{lstlisting}[language=python,mathescape=true]
SliceInternalJoin(
  $\Lambda$,
  feature_sets=[$\calF_1, \calF_2, \dots, \calF_k$],
  join_conditions,
)
\end{lstlisting}
where {\sf join\_statement} is a traditional SQL join statement, computes a slice relations by applying {\sf join\_statement} to the feature tables of each slice tuple.
\end{definition}

\subsection{External operations}

\subsubsection{SliceProject}
\label{sec:slice-project}

In the classic relational algebra, Project can be thought of as a restriction operation, which restricts data using schemas. A function view can clarify this: Recall that a tuple of a relation can be viewed as a function $f(i, \cdot)$ where $i$ is some fixed row index, and the second component is a variable which maps an attribute $a \in \calA$ to some value. Now consider $\calA' \subseteq \calA$, a subset of $\calA$, then we can define the restricted function $f|_{\calA'}$, simply as $f|_{\calA'}(i, a) = f(i, a)$, for any $a \in \calA'$. This $f|_{\calA'}$, the restriction of $f$ to $\calA'$, is the \emph{projection} of the tuple to $\calA'$.

SliceProject generalizes this and \emph{restricts a slice relation using slice-relation schemas}. To this end, recall that we have two kinds of schemas: Schemas of the regions, and schemas of the feature tables, and we can restrict both. To make the discussion more precise, let us first generalize the restriction operation of the tuples to that of the slice tuples.
\begin{definition}[\textbf{Projection of a slice tuple}]
Let $\langle \overline{\Gamma}, \overline{\calF} \rangle$ be a slice relation schema. Let $f(r, \cdot)$ be a slice tuple, where $\schema(r) \in \overline{\Gamma}$, and it maps a feature schema $\calF \in \overline{\calF}$ to a relation with schema $\calF$. Now, suppose we have a subset of region spaces, that is $\overline{\Gamma}' = [\Gamma'_1, \dots, \Gamma'_{m'}] \subseteq \overline{\Gamma}$, and $\overline{\calF}' = [\calF'_1, \calF'_2, \dots, \calF'_{k'}] \subseteq \overline{F}$, then the \emph{projection of the slice tuple} to $\langle \overline{\Gamma}', \overline{\calF}'\rangle$ is as follows: First, if $\schema(r) \notin \overline{\Gamma}'$, the slice tuple is discarded, otherwise, $\schema(r) \in \overline{\Gamma}'$, then we define $f|_{\overline{\calF}'}(r, \cdot)$, simply as $f|_{\overline{\calF}'}(r, \calF) = f(r, \calF)$, for $\calF \in \overline{\calF}'$.
\end{definition}

Our discussion leads to the following definition SliceProject:

\begin{definition}[\textbf{SliceProject}]
\label{def:slice-project}
Let $\Lambda$ be a slice relation with schema $(\overline{\Gamma}, \overline{\calF})$, where $\Gamma=[\Gamma_1, \dots, \Gamma_m]$, and $\overline{\calF} = [\calF_1, \dots, \calF_k]$. Let $\overline{\Gamma}' = [\Gamma'_1, \dots, \Gamma'_{m'}] \subseteq \overline{\Gamma}$ be a subset of region schemas, and $\overline{\calF}' = [\calF'_1, \dots, \calF'_{k'}] \subseteq \overline{\calF}$ be a subset of feature schemas, then the SliceProject operator
\begin{lstlisting}[language=python,mathescape=true]
SliceProject(
  $\Lambda$,
  region_schemas=[$\Gamma'_1, \Gamma'_2, \dots, \Gamma'_{m'}$],
  feature_schemas=[$\calF'_1, \calF'_2, \dots, \calF'_{k'}$],
)
\end{lstlisting}
computes a slice relation (i.e., a set of slice tuples) by projecting each slice tuple in $\Lambda$ to $\langle \overline{\Gamma}', \overline{\calF}' \rangle$.
\end{definition}

For example, suppose that we have a slice-relation $\Lambda$  of schema $\langle \overline{\Gamma}, \overline{\calF} \rangle$, as visualized below,
\begin{table}[htb]
  \centering
  \begin{tabular}{|l|l|l|l|l|}
    \toprule
    Region & $\calF_1$ & $\calF_2$ & $\cdots$ & $\calF_k$\\\hline\hline
    $r_1$ & $X^1_1$ & $X^1_2$ & $\cdots$ & $X^1_k$ \\ \hline
    $r_2$ & $X^2_1$ & $X^2_2$ & $\cdots$ & $X^2_k$ \\ \hline
    ... & ... & ... & ... & ... \\\hline
    \bottomrule
  \end{tabular}
  \label{tab:slice-relation-before-projection}
\end{table}

Then, a slice-project SliceProject($\Lambda$, region\_spaces=[$\Gamma_1$], feature\_schemas=[$\calF_1$]), projects $\Lambda$ to the following:
\begin{table}[htb]
  \centering
  \begin{tabular}{|l|l|}
    \toprule
    Region & $\calF_1$ \\\hline\hline
    $r_1$ & $X^1_1$  \\ \hline
    $r_2$ & $X^2_1$ \\ \hline
    ... & ... \\\hline
    \bottomrule
  \end{tabular}
  \caption{\textbf{Results of projection}. There is only one feature schema now, namely $\calF_1$. All regions have schema $\Gamma_1$.
  }
  \label{tab:slice-relation-after-projection}
\end{table}

\subsubsection{SliceSelect}
\label{sec:slice-select}

In the classic relational algebra, the Select operation applies some predicate to each tuple, and returns the set of tuples where the predicate evaluates to True. Formally, this is written as $\sigma_\phi(X)$ where $\sigma$ denotes the ``selection'' operation, $\phi$ denotes a propositional logic formula, and $X$ denotes an input relation.  In the standard definition, the predicate $\phi$ is a propositional logic formula with several atoms (in the SQL terminology, $\phi$ specifies a WHERE clause), where each item is either

\textbullet\: of the form $\theta(A, c)$, which compares one attribute $A$ with a constant $c$ (e.g., `Revenue > 1000'), or

\textbullet\: of the form $\theta(A, B)$, which involves two two attributes $A$ and $B$ (e.g., `Price > Cost + 5')

For slice relations, the idea is similar, we apply a predicate to each slice tuple, and return the set of slice tuples where the predicate evaluates to True. Formally, we have a predicate $\Phi$, and the selection is denoted as
$\SliceSelect_\Phi(\Lambda)$, where $\Lambda$ is an input slice relation.

However, the predicate becomes more complex. For example, suppose we have a slice tuple of the form $(\text{Region}=r, \calF_1=X_1, \calF_2=X_2)$, then we can have the following different filters:

\textbullet\: we can predicate on the region $r$, which is a tuple. To predicate on regions of certain schema, we use the notation $\text{Region}[\Gamma]$ to refer to regions of schema $\Gamma$. For example the predicate `$\text{Region}[[\text{Device}]].\text{Device} = \text{Pixel}$' picks up regions in the [Device] space and Device takes value Pixel.

\textbullet\: we can apply any scalar-valued function $f$ to convert a feature table to a scalar, and compare it with a constant, for example  $f(\calF_1) > c$. Note that here $f$ can be any transformation that converts a table to a scalar (for example, we can do a summation on a column), so this is really broad.

\textbullet\: we can apply two scalar-valued functions $f$ and $g$, to $X_1$ and $X_2$, and compare the two resulting scalars, for example $f(\calF_1) < g(\calF_2)$.

\textbullet\: we can apply a table-valued function $F$ to convert a feature table to another table, and compare it with a table constant using a set operation (such as $F(\calF_1) \subseteq C$). In particular, $F$ can be any relational query. For example, $\pi_{[\text{Date}]}([\text{Device}, \text{Date}]) \subseteq [(\text{Date=2025-01-01}), (\text{Date=2025-01-02)}]$, which projects the column labeled as [Device, Date] to [Date], and check whether the set of dates is a subset of [2025-01-01, 2025-01-02].

\textbullet\: we can apply two table-valued functions $F, G$ to transform two feature tables, respectively, and compare using a set operation (such as $F(\calF_1) \subseteq F(\calF_2)$). For example, $\pi_{\text{[Date]}}([\text{Device}, \text{Date}]) \subseteq [\text{Date}]$, which projects the column labeled as [Device, Date] to [Date], and check whether the set of dates is a subset of the feature table specified in the column [Date].

Finally, of course, we can still connect these atoms using propositional logic connectives to form a larger formula.

\begin{example}
Suppose we have a feature table column with feature schema $\calF=$[Date, Revenue], then the predicate `MAX([Date, Revenue], metric=Revenue) > 100.0' computes the maximum revenue over dates of the feature table with schema [Date, Revenue], and filters slice tuples where the max daily revenue is bounded by 100.
\end{example}

\subsubsection{SliceTransform}
\label{sec:slice-transform}

Recall earlier that we have the $\mu$ operator in the classic relational algebra, where we mutate column values of each tuple. SliceTransform is the analog of $\mu$ but for slice relations: The idea is that since now we have tables within each slice tuple due to the nesting, we can transform these tables by applying arbitrary table-to-table transformations, which in particular includes arbitrary SQL.  In this section we starts with a special case of this idea, and then generalize it subsequently. Recall that a slice is a $(\region=r, \calF=X)$ pair, which locates a particular cell in a slice relation.

\begin{definition}[\textbf{1-Slice Transformation}]
\label{def:1-slice-transformation}
A 1-slice transformation is a function that transforms a slice $(r, X)$ to a table $X'$, such that $\schema(X')$ only depends on $\schema(X)$.
\end{definition}

Given a slice transformation $\tau$, the attributes it references is denoted as $\schema(\tau)$.

\begin{definition}[\textbf{SliceTransform}]
\label{def:slice-transform}
Let $\Lambda$ be a slice-relation, $\calF^{(1)}, \dots, \calF^{(\ell)}$ a sequence of feature schemas, and $\tau_1, \dots, \tau_\ell$ be a sequence of slice transformations. Suppose each slice transformation has output schema $\calF'_{(i)}$. We assume the set of output schemas, plus the slice columns that are not designated form a legitimate slice relation schema. Then a SliceTransform operator
\begin{lstlisting}[language=python,mathescape=true]
SliceTransform(
  $\Lambda$,
  slice_transformations=[$\left\langle \calF_{(1)}, \tau_1 \right\rangle, \dots, \left\langle \calF_{(\ell)}, \tau_\ell\right\rangle$],
  dimensions=dimensions,
)
\end{lstlisting}
where for each slice transformation, $\schema(\tau_i) \subseteq \calF_{(i)}$, computes an output slice-relation as follows:

\textbf{(1) Applying slice transformations to slice tuples}. For each slice tuple in $\Lambda$, say $\langle \region=r: \calF_1=X_1, \dots, \calF_k=X_k\rangle$. We then apply the slice transformation $\tau_i$ to the feature table of schema $\calF_{(i)}$, projected to $\schema(\tau_i)$, giving a result tables $X'_{(i)}$ with schema $\calF'_{(i)}$. All the other slice columns not listed remain intact.

\textbf{(2) Forming a slice relation}. All slice tuples are collected into an output slice relation, and dimensions is set to the argument {\sf dimensions}.
\end{definition}

\subsubsection{SliceJoin}
\label{sec:slice-join}


\begin{definition}[\textbf{SliceJoin}]
\label{def:slice-join}
SliceJoin is an operator
\begin{lstlisting}[language=python,mathescape=true]
SliceJoin(
  $\Lambda_1$, $\Lambda_2$,
  region_join_conditions,
  dimensions=dimensions,
)
\end{lstlisting}
which computes an output multi-relation as follows:

\textbf{(1)} \textbf{Join slice tuples.} Consider a typical slice tuple $\langle r: \calF_1=X_1, \dots, \calF_k=X_k \rangle$ from $\Lambda_1$, and another typical slice tuple $\langle r': \calF_1'=X_1', \dots, \calF_{l}'=X'_{l} \rangle$ from $\Lambda_2$. These two slice tuples are joinable if the regions $r$ and $r'$ satisfy the `region\_join\_conditions'. If joinable, they are joined as $\widetilde{r}: \calF_1=X_1, \dots, \calF_k=X_k, \calF_1'=X_1', \dots, \calF_l'=X_l' \rangle$, where $\widetilde{r}$ is formed by concatenating the two regions $r, r'$, with appropriate renaming of region attributes (same as join in the classic relational algebra..

\textbf{(2) Repeat for all joinable slice tuples.} The above is repeated for any two joinable tuples. The resulting set of slice tuples form the result slice relation.

Same as join in the relational algebra, proper renaming of slice columns will be performed if needed. For example if both slice relations have [Date, Cpc], then the renaming is [Date\_l, Cpc\_l] for the slice column on the left, and [Date\_r, Cpc\_r] for the slice column on the right.
\end{definition}

For {\sf region\_join\_conditions}, we consider two conditions, one is that $r = r'$ (that is, equi-join on regions), and second, we consider $r \prec r'$, defined as follows:
\begin{definition}[\textbf{Region partial order}]
\label{def:region-partial-order}
For two regions $\gamma_1$ and $\gamma_2$, $\gamma_1 \prec \gamma_2$, if every attribute value in $\gamma_2$ is also in $\gamma_1$. 
\end{definition}

In other words, $\gamma_1$ is more fine-grained, while $\gamma_2$ is more coarse-grained (e.g., (State=CA, City=MTV) $\prec$ (State=CA)).

\subsection{Further extensions}
\subsubsection{SliceTransform with reference features}
\label{sec:slicetransform-with-ref}
We first define reference feature tables
\begin{definition}[\textbf{Reference feature tables}]
\label{def:ref-features}
Given a slice-relation $\Lambda$, and a feature set $\calF$, the reference feature table w.r.t. $\calF$ is the slice indexed by $\calF$ and the empty region tuple.
\end{definition}

The following version of SliceTransform, which uses a 2-slice transformation so as to compare a slice at a region with the reference slice at the region (), is very useful for our applications.
\begin{definition}[\textbf{SliceTransform with reference features}]
\label{def:region-transform-ref}
It is the same as SliceTransform except that we use slice transformations such that, given one feature set $\calF$, analyze the two input slices $(\region=r, \calF=X)$ and $(\region=(), \calF=X_0$).
\end{definition}

Slice transformations with reference features significantly broaden the analytical problems one can express in one SliceTransform.

\section{Operations on relation spaces}
\label{sec:operations-rs}
In many scenarios, analysts are not interested in explicitly invoking the {\sf Represent} operator; 
instead, they simply want to transform a relation space directly. 
To support such workflows, we introduce a set of high-level operations on relation spaces, 
which internally compose multiple slice relation operations. 
These operators serve as convenient, end-to-end abstractions—effectively acting as 
``mega operators'' tailored for common use cases. 
Conceptually, each such operator follows a three-step pattern: 
it first uses {\sf Represent} to convert the input relation space into a slice relation, 
then applies one or more transformations on the slice relation, 
and finally flattens the result back into a relation space. 
This process can be visualized as:

The following diagram summarizes the pattern
\begin{align*}
\text{relation space} 
\xrightarrow{\text{Represent}} \text{slice relation}
\xrightarrow{\text{Transformations}} \text{output slice relation}
\xrightarrow{\text{Flatten}}  \text{output relation space}
\end{align*}

An important such operator is Crawl, which is defined as follows:
\begin{definition}[\textbf{Crawl}]
\label{def:crawl}

Let $\Psi$ be a relation space. The {\sf Crawl} operator:

\begin{lstlisting}[language=python,mathescape=true]
Crawl(
  $\Psi$,
  region_schemas,
  slice_transformations,
  predicates,
  dimensions,
)
\end{lstlisting}

performs the following sequence of operations:

\begin{enumerate}
\item Let {\sf feature\_schemas} be the set of feature schemas required by {\sf slice\_transformations}. Construct a slice relation via:
\begin{lstlisting}[language=python,mathescape=true]
S1 = Represent(
  $\Psi$,
  region_schemas,
  feature_schemas,
)
\end{lstlisting}

\item Apply the specified slice transformations:
\begin{lstlisting}[language=python,mathescape=true]
S2 = SliceTransform(
  S1,
  region_schemas,
  slice_transformations,
)
\end{lstlisting}

\item Filter the resulting slice tuples using the given predicates:
\begin{lstlisting}[language=python,mathescape=true]
S3 = SliceSelect(
  S2,
  predicates,
)
\end{lstlisting}

\item Flatten the selected slice relation back into a relation space:
\begin{lstlisting}[language=python,mathescape=true]
$\Psi'$ = Flatten(
  S3,
  dimensions,
)
\end{lstlisting}
\end{enumerate}
\end{definition}

\section{Our current implementation}
\label{sec:impl}
We describe our current implementation of MRA at Google, 
focusing on the \textsf{Crawl} operator for simplicity. 
The implementation is built on top of \textsf{F1 Query}, 
Google's internal distributed relational query engine.

A core aspect of our approach is the utilization of a unique feature 
within \textsf{F1 Query}'s \textsf{Table-Valued Functions (TVFs)}: 
the \textsf{@partition\_by} directive. This directive, not generally 
available in other major query engines like \textsf{Postgres} or \textsf{BigQuery}, 
allows us to implement a process that is conceptually similar to columnar processing, 
but which operates at the level of slices rather than individual tuples.

The operational flow begins with an input relation space. 
For each slice transformation and region space defined in the query, 
the system matches the appropriate source table. This table is then partitioned, 
creating a collection of slices upon which the specified slice transformation is evaluated. 
After this process is repeated for all required transformations and region spaces, 
the resulting processed slices are aligned by their common regions to form slice tuples. 
These tuples are then filtered according to the query's predicates. 
As a final step, the filtered slice relation is flattened back into a 
standard relation space to produce the final output.

While this architecture is powerful, our large-scale analytical workloads 
generated performance challenges for the \textsf{@partition\_by} directive, 
which was not originally intended for this type of intensive use. 
To overcome these bottlenecks, we designed and implemented a specialized execution engine, 
termed the \textsf{Region Evaluation System} (\textsf{RES}). 
The \textsf{RES} is purpose-built to manage the distribution, 
scheduling, and processing of slices at production scale. Unlike a generic 
partitioning mechanism, \textsf{RES} incorporates a sophisticated, workload-aware scheduler 
that understands the hierarchical relationships between region spaces. This awareness is key
to enabling advanced optimizations, as we describe next.

\subsection{Optimization Strategies}

While a naive execution of a \textsf{Crawl} operation may appear brute-force, 
the evaluations of slice transformations across different region spaces are often 
highly correlated. The \textsf{RES} is designed to exploit these correlations. 
To illustrate, we consider optimizations based on the Apriori property, 
which relies on the natural partial order that exists among regions.

\begin{definition}[\textbf{Region Lattice}]
\label{def:region-lattice}
For two regions $\gamma_1$ and $\gamma_2$, we say $\gamma_1 \prec \gamma_2$ if every 
attribute-value pair in $\gamma_2$ is also present in $\gamma_1$. In other words, 
$\gamma_1$ is more fine-grained, while $\gamma_2$ is more coarse-grained 
(e.g., \textsf{(State=CA, City=MTV) $\prec$ (State=CA)}).
\end{definition}

This region lattice allows us to define a monotonicity property for metrics 
computed by slice transformations.

\begin{definition}[\textbf{Apriori Property of Slice Transformation Results}]
\label{def:apriori}
Let $\sigma$ be a metric computed by a slice transformation. The metric is said 
to satisfy the Apriori property if for any two regions where $\gamma_1 \prec \gamma_2$, 
it holds that $\sigma[\gamma_1] \prec \sigma[\gamma_2]$\footnote{$\prec$ denotes a partial order 
which may not be the linear order $<$ on real numbers.}. Here, $\sigma[\gamma]$ denotes 
the value of metric $\sigma$ on region $\gamma$.
\end{definition}

\textbf{Apriori Property and Early Stopping.}
If a slice transformation computes a metric that satisfies the Apriori property, 
filtering on that metric enables powerful "early stopping" optimizations. 
For example, if a coarse-grained region fails to meet a filter predicate, 
all finer-grained regions nested within it can be pruned from evaluation, 
as they are guaranteed to fail as well.

\textbf{Degree-First Evaluation.}
This principle can be implemented using various evaluation strategies. A \emph{degree}-first%
\footnote{The number of attributes in a region space is called its \emph{degree}.} 
evaluation, for example, processes all degree-1 region spaces (e.g., \textsf{[Device]}, 
\textsf{[Browser]}) before moving to degree-2 spaces (e.g., \textsf{[Device, Browser]}). 
By filtering results at each degree, the system can dramatically reduce the number of 
candidate regions for higher-degree spaces. This strategy is equivalent to a breadth-first 
search over the region lattice and effectively recovers the well-known Apriori algorithm 
for frequent itemset mining~\cite{AprioriAlgorithm} as a special case of our generalized 
optimization framework.

\textbf{Beyond Degree-First Evaluation.}
Other strategies are also possible within the \textsf{RES} framework, such as 
depth-first evaluation or an \emph{optimistic region evaluation}. The latter evaluates 
all region spaces in parallel but performs a cheap check using the Apriori property 
to determine if a costly slice transformation can be skipped for a given region. 
The optimal choice of strategy is a complex problem that depends on the data 
distribution and query structure; a detailed exploration is beyond the scope of 
this paper and is an area of future work.

\section{Applications}
\label{sec:applications}
We survey current applications of MRA.

\textbf{\em Application I: System Monitoring}.
Many of our clients operate complex systems, making it critical to monitor their health effectively.
This monitoring often requires a fine-grained understanding of specific aspects of the system,
such as the gradual increase in average latency for a particular version in a specific geographic region.
Example analytical questions include:

\textbf{(i)} \textit{Which slices exhibit time series anomalies of sudden spikes?}
This requires slice transformations with advanced statistics, such as CausalImpact analysis~\cite{BGKRS15}.

\textbf{(ii)} \textit{Which slices show a slow deterioration in key performance indicators (KPIs) over time?} 

\textbf{(iii)} \textit{Which slices have forecasts that indicate a significantly increased cost of operation, necessitating early preparations?}

\textbf{\em Application II: Business Intelligence}.
Several of our major Business Intelligence platforms need to generate data insights that help users understand current trends and identify new business opportunities.
The requirements are often complex and ad-hoc, such as analyzing which campaigns and ad groups are influencing top-level revenue changes
and how they correlate with budget allocation. Example analytical questions include:

{\bf (i)} {\em Which slices contribute significantly to a top-level metric change?} 
When the metric is summable, attributing change to individual slices is straightforward. 
However, for non-summable metrics, the contribution problem becomes more complex. 
To address this, we developed a slice transformation that incorporates a reference feature, 
drawing on the classic Aumann-Shapley method~\cite{AS2015}. 
This approach applies to any metric that admits a differentiable structure. 
In Section~\ref{sec:aumann-shapley-sf}, we derive the contribution formula 
for a class of density metrics—metrics where both the numerator and denominator 
can be aggregated using {\sf SUM}.

{\bf (ii)} {\em Which slices reveal interesting correlations between key metrics?} 
While correlation is a fundamental concept in statistics, 
classical methods often prove ineffective in practice, 
as they typically rely on idealized assumptions—such as linearity or normality—that rarely hold in real-world data. 
To address this, we developed new correlation techniques that are more robust and practically useful. 
One such method is the {\em cross-rank correlation}, which we describe in Section~\ref{sec:cross-corr-sf}. 
When used as slice transformations within a multi-relational operation, 
these analyses become especially powerful: 
slicing narrows the analysis to smaller, entity-aligned subsets of data, 
where correlations tend to be more localized, more interpretable, 
and more statistically detectable. 
This makes the slice granularity a natural and effective unit for discovering meaningful metric interactions.

\textbf{(iii)} \textit{Which slices differ significantly from a reference slice, indicating potential business opportunities?} 

\textbf{(iv)} \textit{Which slices have forecasts that suggest a significant decrease in revenue, prompting a closer examination of that subpopulation?}

\textbf{\em Application III: Experimentation Analysis}.
Several experimentation applications have been developed to slice data for A/B testing,
allowing for a detailed understanding of an experiment's impact. Example analytical questions include:

\textbf{(i)} \textit{Which slices are significantly overperforming in the experiment, indicating potential business opportunities?}

\textbf{(ii)} \textit{Which slices are significantly underperforming, suggesting they should be isolated for further investigation?}

\begin{figure}[htb]
\centering
\includegraphics[width=.35\linewidth]{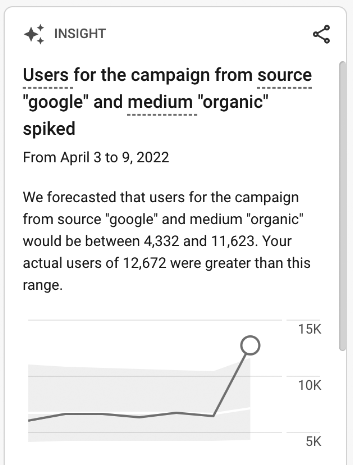}
\includegraphics[width=.35\linewidth]{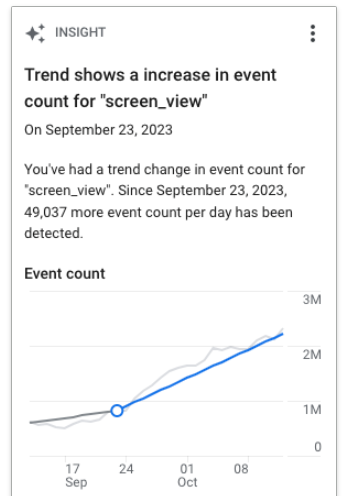}
\cprotect\caption{Our multi-relational service powers two Google Analytics insights, Contributor~\cite{GAContributionInsights} and Trend-Change~\cite{GATrendChangeInsights}. Each highlights interesting slices, such as the one on the left with region (source=`google', medium=`organic').
}
\label{fig:ga-insights}
\end{figure}

Multi-relational abstractions benefit all these applications:

\textbf{Novel or Enhanced Functionalities}. Our multi-relational analytics service has 
enabled the development of novel BI functionalities.
Externally, Google Analytics trend change insights~\cite{GATrendChangeInsights}
and contribution insights~\cite{GAContributionInsights} are based on our service; see Figure~\ref{fig:ga-insights}.
Snowflake Top Insights~\cite{SnowflakeTopInsights} is MultiSelect with a simple slice transformation.
Internal at Google, BI tools built using our service have identified revenue 
anomalies and their correlation with campaign budgets, leading to revenue increases. 
These BI tools represent advancements  beyond the capabilities of pre-existing tools.

\textbf{Improved Performance and Reduced Costs.} The unified logical design of multi-relational 
algebra enables systematic optimizations across diverse analytical tasks. This results in 
improved latency, scalability, and often reduced resource costs when migrating from ad-hoc 
systems to a multi-relational analytics service.

\textbf{Migrating to a Relational Ecosystem}. Experimentation analysis platforms often 
employ custom slicing-analysis kernels using Apache Beam~\cite{PythonBeam} due to its superior flexibility, 
but these can be complex, involving tens of thousands of lines of code. Multi-relational 
algebra pipes streamline these into concise pipelines, improving interoperability with 
SQL-based systems and simplifying both experimentation analysis and integration with 
relational databases.

\section{Related Work}
\label{sec:related}


{\bf Nested Relational Model.} 
The nested relational model and its associated algebra~\cite{NestedRelation} 
extend the classical relational model by allowing attributes to take entire relations as values. 
Compared to {\sf SliceRelation}, this model permits arbitrary levels of nesting, 
which can quickly become difficult for humans to reason about. 
MRA explicitly prohibits such arbitrary nesting: 
a {\sf SliceRelation} is not a relation in the nested relational sense, 
nor should it be viewed as a restricted instance of one. 
In MRA, we support only one level of nesting, and the design reflects this constraint. 
Each {\em slice tuple} is indexed by a {\em region}, 
and each column in a {\sf SliceRelation} uniformly holds a relation-valued feature table. 
This structural regularity enables predictable, analyzable semantics 
that differ from the general-purpose but less disciplined nesting allowed in the classic model.

{\bf OLAP Cube.} 
OLAP cubes~\cite{OLAPCube} organize data into multi-dimensional arrays and support navigation and aggregation 
over predefined dimensions. Query languages like MDX~\cite{MDX} allow users to slice and dice the data interactively, 
retrieving specific aggregates such as ``yearly revenue by device from 2014 to 2024.'' 
While MRA’s concept of a ``slice'' is superficially similar, its purpose and structure are fundamentally different. 
OLAP systems focus on exploring aggregations through manual, navigational queries, 
whereas MRA provides a declarative algebra for constructing, transforming, and analyzing 
collections of slice tuples—each containing relation-valued features—at multiple granularities. 
Rather than retrieving summaries, MRA operators are designed to evaluate, compare, 
and select slices based on analytical criteria such as anomalies or metric contributions.

{\bf Group-by GROUPING SETS and CUBE.} 
Group-by aggregation is a fundamental analytical operation in SQL. 
A query like `SELECT A, SUM(B) FROM X GROUP BY A` partitions table $X$ by values of attribute $A$, 
sums $B$ within each group, and returns a table of aggregated results. 
{\sf GROUPING SETS} generalizes this idea to compute multiple group-bys in a single query. 
For example, `SELECT A, SUM(B) FROM X GROUP BY GROUPING SETS ((A), ())` computes both 
the group-by on $A$ and the global aggregation (i.e., `SELECT SUM(B) FROM X`). 
Logically, this yields two separate tables with schemas [A, B] and [B], 
but SQL flattens them into a single output by using NULL padding, 
resulting in tuples like (A = NULL, B = ...). 
To simplify specification of many groupings, SQL includes syntactic extensions such as 
{\sf GROUP BY CUBE}~\cite{GrayBLP96} and {\sf GROUP BY ROLLUP}~\cite{GroupByRollup}. 
For instance, `GROUP BY CUBE(A)` automatically includes both grouping sets (A) and (), 
and more generally, CUBE over $n$ attributes produces all $2^n$ possible subsets. 

While useful for expressing aggregations at different granularities, 
these constructs lack explicit modeling of schema-level structure. 
In contrast, MRA introduces {\sf RelationSpace} as a logical abstraction 
that retains the identity of each grouping explicitly, avoids NULL padding, 
and supports downstream operations such as per-entity feature evaluation and transformation.

{\bf User-Defined Aggregation (UDA).} 
User-Defined Aggregates (UDAs), such as those supported in PostgreSQL~\cite{PostgreSqlUDA}, 
enable customized aggregation logic, allowing complex computations to be performed within each group. 
While UDAs extend the expressiveness of standard SQL aggregation, they are typically scoped to single-column processing, 
making it difficult to operate over multiple attributes simultaneously. 
Some systems address this by packing multiple columns into array- or JSON-valued fields, 
but this approach introduces encoding complexity and sacrifices schema clarity. 
UDAs generally follow a state-transition model: an initial state is established, 
and updated incrementally as tuples are processed. 
While this model is powerful for certain aggregation tasks, 
it lacks structural support for organizing and transforming multi-table or multi-feature inputs. 
In contrast, MRA offers explicit data abstractions—{\sf RelationSpace} and {\sf SliceRelation}—that 
treat sets of relations as first-class objects and support feature-level analysis and transformation 
at varying granularities.




{\bf DIFF.} 
The $\sf DIFF$ operator~\cite{AKSGXSASM21} is a relational aggregation construct designed for explanation-oriented queries. 
It generalizes several prior data explanation systems, including Scorpion~\cite{WM13} and Data X-Ray~\cite{WDM15}, 
and provides a declarative interface for identifying attributes that contribute to differences between two input tables. 
A typical usage takes the form:
\begin{lstlisting}[language=sql,mathescape=true]
Table T1 DIFF Table T2
ON A, B, C
COMPARED BY diff_metric >= threshold
MAX ORDER n
\end{lstlisting}
Here, $\sf diff\_metric$ refers to one of several predefined statistical measures 
(e.g., $\sf support$, $\sf odds\_ratio$, $\sf risk\_ratio$, $\sf mean\_shift$) used to evaluate the strength of difference 
between the two datasets $T1$ and $T2$ along dimensions $A$, $B$, and $C$. 
The $\sf MAX\ ORDER\ n$ parameter restricts the explanations to combinations of at most $n$ attributes. 
Unlike MRA, which offers a general-purpose algebra for constructing multi-granular analytical pipelines, 
$\sf DIFF$ is tailored to a specific form of comparison analysis, and does not support relational composition or feature-level transformations.

{\bf AutoSlicer.} 
AutoSlicer~\cite{LRS22} is an automated slicing system designed to identify subsets of evaluation data 
where a trained model exhibits anomalous behavior. 
It relies on a candidate slice generator that explores a search space of selection predicates 
to define slices—e.g., $\verb|WHERE A = a|$ or $\verb|WHERE A = a AND B = b|$, 
where $A$ and $B$ are attributes and $a$, $b$ are specific values. 
These slices are then ranked according to metrics that quantify model performance degradation. 
AutoSlicer is implemented in Apache Beam~\cite{PythonBeam}, 
which provides a more flexible data model and transformation primitives than traditional relational algebra. 
Unlike MRA, which offers a composable algebra for constructing and analyzing relation-valued feature slices, 
AutoSlicer focuses on predicate-based discovery over flat tables and does not model multi-relational or multi-granular structure.

\section{Conclusion}
\label{sec:conclusion}
This paper introduces Multi-Relational Algebra (MRA), 
a principled extension of classical relational algebra 
designed to support multi-granular analytics through two new data abstractions: 
{\sf RelationSpace} and {\sf SliceRelation}. 
These abstractions enable analysts to reason about collections of relations 
and relation-valued features in a structured and composable way. 
MRA draws inspiration from OLAP cubes and nested relational models, 
but generalizes these ideas into a unified framework that remains compatible 
with the relational model. 
We implemented MRA as a DSL exposed via Table-Valued Functions, 
and deployed it as a production service within Google, 
where it has seen broad adoption across teams and product areas. 
Looking ahead, we plan to extend MRA with new optimization techniques and deeper SQL integration to support increasingly complex analytical workflows on a solid algebraic foundation.

{\small
\bibliographystyle{alpha}
\bibliography{paper}
}

\appendix
\section{Example slice transformations and compositions}
\label{sec:mr-analytics}
Multi-relational operators with slice transformations capture a wide-variety of problems, significantly broadening the scope of previous work, which are limited to simple descriptive statistics. This power stems from two aforementioned characteristics: \textbf{(1)} the information unit of the algebra is a slice tuple $(r, X_1, \dots, X_k)$, which allows users to concisely express the need for analyzing feature sets aggregated at different granularity. and \textbf{(2)} multi-relational operators offer powerful transformations of sets of slices tuples, and can be easily composed for more complex tasks. As initial illustrations, MultiSelect combined with simple slice transformations can easily express the functionality of two prior works, DIFF~\cite{ABDGMNRS18} and AutoSlicer~\cite{LRS22}.

{\textbullet}\ \textbf{DIFF (Abuzaid et al. (2021))}.
To cover DIFF, we can encode the two  input tables as a single table with a True/False column, `is\_test', to identify them. The statistics can be wrapped into a slice transformation that loads the `is\_test' column as a feature. This automatically covers special cases of the DIFF operator, such as Scorpion~\cite{WM13} and Data X-Ray~\cite{WDM15}.

{\textbullet}\ \textbf{AutoSlicer (Liu et al. 2022)}.
\cite{LRS22} To cover AutoSlicer, its ``search space'' is exactly the region space and the analysis on slices can be easily captured by a 1-slice transformation (Definition~\ref{def:1-slice-transformation}).

In the following we give a few more examples where, interestingly, multi-relational analytics also reveal novel analytical problems.


\subsection{Aumann-Shapley Slice Transformation}
\label{sec:aumann-shapley-sf}
A common data insight problem is the \emph{region attribution problem}: let $\mu$ be a metric, and the data is divided into test and reference segments. For the test segment, we use $\mu_t$ to denote the metric value of the entire test segment. For the reference segment, we use $\mu_c$ to denote the metric value of the entire reference segment. For a region $\gamma$, we use $\mu_t[\gamma]$ to denote the metric value of the region $\gamma$ in the test segment, and use $\mu_c[\gamma]$ to denote the metric value of the region $\gamma$ in the reference segment. With this setup, we have a \emph{population metric change} $\Delta = \mu_t - \mu_c$. Conceptually, the region attribution problem asks ``\emph{how much does a region $\gamma$ contribute to the change $\Delta$?}'' Formally, we have the following definition.

\begin{definition}[\textbf{Region Attribution Problem}]
\label{def:region-attribution-problem}
In the Region Attribution Problem, the goal is to find a \emph{region attribution} $\ras(\cdot)$,
which is a function that assigns a score to each region, that satisfies:

\textbf{(1) (Completeness)} The empty region, corresponding to the entire population, 
satisfies $\ras(()) = \Delta$.

\textbf{(2) (Additivity)} If $\gamma$ is a disjoint union of $A$ and $B$, then 
$\ras(\gamma) = \ras(A) + \ras(B)$. This is to say that the attribution is 
consistent across different decompositions of a region.
\end{definition}

This problem is trivial if the metric summable (e.g., Revenue), but becomes quite non-trivial if the metric is \emph{non-summable}. Interestingly, we found no previous work has systematically studied this seemingly basic problem, and we present the Aumann-Shapley slice transformation, a \emph{2-slice transformation} which is an application of the Aumann-Shapley method (\cite{SS11,AS2015}), as a principled solution to this problem. To describe this slice transformation, we need the concept of a \emph{Region-Ambient Metric Model}: A region-ambient metric metric model encodes how to combine metrics from a region $\gamma$, and metrics of the ambient of the region $\overline{\gamma}$ (i.e., data \emph{not} in $\gamma$), to recover the \emph{population} metric value. The metrics needed in a metric model may be more than $\mu$ itself. We give two examples of metric models:

\textbf{(i)} \textbf{A summable metric} (e.g., Revenue).
In this case, a region-ambient metric model is simply $F(w, \overline{w}) = w + \overline{w}$.
This metric model uses two variables: $w$, representing the metric value of a specific region,
and $\overline{w}$, representing the metric value of the ambient.

\textbf{(ii)} \textbf{A density metric} (e.g., CostPerClick).
$\mu$ is a density metric if $\mu=W/S$ where both $W$ and $S$ are summable.
A natural metric model for density has parameters:
$F(w, \overline{w}, s, \overline{s}) = (w+\overline{w})/(s+\overline{s})$,
where $w, s$ represent numerator and denominator values of a region,
and $\overline{w}, \overline{s}$ represents numerator and denominator of the ambient. 

Let $F$ be a metric model with parameters $z_1, z_2, \dots, z_n$.
With the test/reference split, we thus have two states of this metric model.
In the reference period, the parameters take on the values $z_1^c, z_2^c, \dots, z_n^c$,
which are collectively represented by the vector $P_0=(z_1^c, z_2^c, \dots, z_n^c)$.
Similarly, in the test period, the parameter values are into another vector $P_1=(z_1^t, z_2^t, \dots, z_n^t)$.
Now, to apply the Aumann-Shapley method, we consider a line path connecting these two points:
$h(t) = (1-t)P_0 + tP_1, t \in [0,1]$.
Then, for $G(t) = F(h(t))$,
\begin{align*}
  \Delta & = F(P_1) - F(P_0)
  = G(1) - G(0) 
  = \int_0^1 G'(t) dt \\
  & = \int_0^1 \sum_{i=1}^n
  \frac{\partial F}{\partial z_i}\frac{\partial z_i}{\partial t}dt 
  = \sum_{i=1}^n \left(\int_0^1
  \frac{\partial F}{\partial z_i}\frac{\partial z_i}{\partial t}dt\right)
\end{align*}
Therefore, for the attribution to $z_i$, we can define it as
$\ras(\gamma; z_i)=\int_0^1 \frac{\partial F}{\partial z_i}\frac{\partial z_i}{\partial t} dt$.
Then we can define the region attribution as the sum of $\ras(\gamma; z_i)$ for $z_i$'s
that represent metric values of the region.
We show this for both summable and density metrics below.

\textbf{Summable metric}.
For a summable metric with metric model $F(w,\overline{w}) = w+\overline{w}$,
$z_1=w$, and $z_2=\overline{w}$.
Since only the variable $w$, represents the value of the region,
the attribution is expressed as $\ras(\gamma) = \ras(\gamma; w)$.
Given the reference and test period parameter vectors
$P_0 = (\mu_c[\gamma], \mu_c - \mu_c[\gamma])$ and
$P_1 = (\mu_t[\gamma], \mu_t - \mu_t[\gamma])$, respectively,
we have
$\ras(\gamma) = \ras(\gamma; w) = \mu_t[\gamma] - \mu_c[\gamma]$,
which recovers the intuitive attribution for summable metrics!

\textbf{Density metric}. For a density metric with metric model \newline
$F(w, \overline{w}, s, \overline{s}) = (w+\overline{w})/(s+\overline{s})$,
both parameter $w$ and parameters $s$ represent metric values of the region,
and so $\ras(\gamma) = \ras(\gamma; w) + \ras(\gamma; s)$.
Given the reference and test period parameter vectors
$P_0=(w_c[\gamma], s_c[\gamma], w_c-w_c[\gamma], s_c-s_c[\gamma])$,
and
$P_1=(w_t[\gamma], s_t[\gamma], w_t-w_t[\gamma], s_t-s_t[\gamma])$,
we have:
\begin{align}
  \label{eq:density-region-attribution}
  \begin{split}
    \ras(\gamma)
    = &\ras(\gamma; w) + \ras(\gamma; s) \\
    = & \frac{\ln(s_t) - \ln(s_c)}{(s_t-s_c)^2} C_\gamma 
      + \frac{s_t[\gamma]-s_c[\gamma]}{s_t-s_c} \Delta
  \end{split}
\end{align}
where $C_\gamma = (w_t[\gamma]-w_c[\gamma])(s_t-s_c) - (w_t-w_c)(s_t[\gamma]-s_c[\gamma])$,
and $\Delta=\mu_t - \mu_c$ is the population metric change.
This formula is much less intuitive compared to the summable case.
Note that reference metrics play an essential role in the formula.
One can prove that the above attribution satisfies both of the
completeness and additivity requirements in Definition~\ref{def:region-attribution-problem}.
However, for complex metric models, the closed forms of integration are harder to derive,
and we use JAX~\cite{jax_2018_github} to compute the integration numerically.

\subsection{Cross-Correlation Slice Transformation}
\label{sec:cross-corr-sf}

While the Aumann-Shapley method can explain metric changes with a differentiable metric model, such a model is not always available. In such cases, rank correlation~\cite{RankCorrelation} can be used to  study their statistical relationship. However, correlation at the aggregate level can be weak and uninformative. This is where MultiSelect proves valuable, enabling data slicing into smaller regions and facilitating the analysis of correlation at a regional level, potentially uncovering more meaningful insights.

\begin{figure}[htb]
\centering
\centering
\includegraphics[width=.8\linewidth]{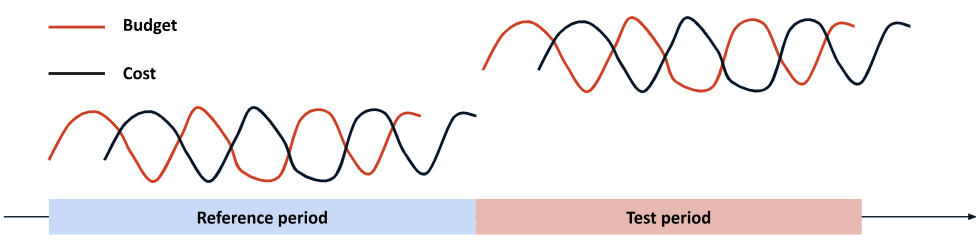}
\caption{\textbf{Cross Rank Correlation:} 
Cost and Budget exhibit perfect negative  correlation within each period (reference or test),
but perfect positive correlation across the two periods.
}
\label{fig:cross-rank-correlation}
\end{figure}

Interestingly, in a test/reference split scenario, which is typical for our use cases, classic notions of rank correlations do not work well, even at the slice level. Figure~\ref{fig:cross-rank-correlation} illustrates a key issue. Here, a budget change at the end of the reference period shows that increasing the budget also increases the cost (positive correlation). However, within each period, Cost and Budget exhibit negative correlation. This mix of positive and negative correlations leads to a weak overall correlation using the classic rank correlation formula, as it considers data from both periods simultaneously. To address this, we propose a novel correlation notion which only pairs of points with one point from each period.  Formally, the test period has $N$ (Budget, Cost) data points, denoted as $U = \{(B^t_i, R^t_i)\}_{i=1}^N$, while the reference period has $M$ data points, denoted as $V = \{B^c_j, R^c_j\}_{j=1}^M$. The cross rank correlation formula is presented in Equation~\ref{eq:cross-rank-correlation}.

\begin{align}
  \label{eq:cross-rank-correlation}
  \text{CrossRankCorr}(U, V) = \frac{\sum_{i \in [N], j \in [M]}\text{sgn}\left(\Big(R^t_i-R^c_j\Big)\Big(B^t_i - B^c_j\Big)\right)}{NM}
\end{align}



\section{More on implementing MRA}

\subsection{Schema type}
\label{sec:schema-type}

Until now, we have been using a set of attributes to refer to a schema. A more rigorous way is to have types for schemas. For example we can define DateCpcSchema for the schema [Date, Cpc], and then we can refer to this type using DateCpcSchema. With this, then a $\represent$ operation of region schema [Device] and feature schema [Date, Cpc] can be specified as

\begin{lstlisting}[language=python,mathescape=true]
Represent(
  $\Psi$,
  region_schema=[DeviceSchema,],
  feature_schema=[DateCpcSchema,], 
)
\end{lstlisting}

\subsection{Schema flexibility}
\label{sec:schema-flexibility}

Note that while a slice relation has a schema, we can easily augment the schema, since we allow it to store data of different schemas.

\section{Data modeling versatility}
\label{sec:versatility}
We note that relation space and slice relation provide a versatile data modeling.

\subsection{SELECT RESULTDB}

A recent work proposed to extend SQL to compute multiple tables as output, terms {\sf SELECT RESULTDB}. As we have discussed earlier in Example~\ref{example:resultdb}, this extension is a special case of SliceInternalSelect applied to a relation space.

\subsection{BigQuery partitioned table}

BigQuery supports an abstraction called {\sf PartitionedTable}~\cite{PartitionedTable}, an example of which is shown in the following Listing~\ref{lst:partitioned-table}.
\begin{lstlisting}[language=SQL,mathescape=true,label={lst:partitioned-table}]
CREATE TABLE
  mydataset.newtable (transaction_id INT64, transaction_date DATE)
PARTITION BY
  transaction_date
AS (
  SELECT
    transaction_id, transaction_date
  FROM
    mydataset.mytable
);
\end{lstlisting}

\begin{figure}[htb]
\centering
\includegraphics[width=0.35\linewidth]{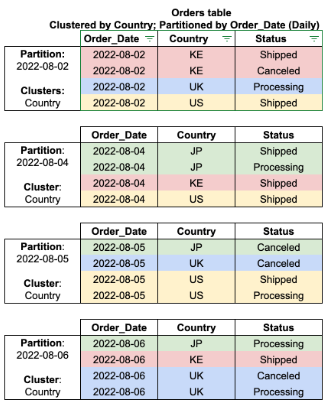}
\caption{Logical representation of a partitioned table. The partition key for each partition, {\sf transaction\_date}, is stored only once per partition.}
\label{fig:partitioned-table-vis}
\end{figure}

A partitioned table is divided into segments, called partitions, with the goal of improving query performance and control costs by reducing the number of bytes read by a query. In the example, one partitions the data by {\sf transaction\_date}. Notably, the representation of a partitioned table is indeed to have one {\sf transaction\_date} per partition, Figure~\ref{fig:partitioned-table-vis} shows the representation of a partitioned table from the BigQuery documentation. The representation is used internally within the BigQuery engine for various purposes. However, to the users who query the partitioned table, the logical representation is still a usual relational table, which indicates logical redundancies of partition keys. For example, if we do {\sf SELECT * FROM mydataset.newtable}, we create multiple duplicated {\sf transaction\_id}.

{\sf PartitionedTable} can be captured precisely by a slice relation by treating partition keys as regions. In particular, the above {\sf PARTITION BY} statement corresponds the following {\sf CREATE SLICE TABLE} operation:
\begin{lstlisting}[language=python,mathescape=true]
CREATE SLICE TABLE SliceTable(
  region_schemas=[[transaction_date],],
  feature_schemas=[[transaction_id],],
)
\end{lstlisting}

Compared to the engine-specific constructs, the MRA enjoys the following benefits:
\begin{itemize}
    \item MRA provides a portable data modeling that is engine agnostic.
    \item We can return slice table directly, eliminating logical redundancies.
    \item More powerful operations: Beyond SQL queries over a partitioned table, MRA provides more powerful operators to the users to explicitly manipulate a partitioned table. This standardization ensures that operations can be performed uniformly and reliably across different data engines, promoting portability and ease of use.
\end{itemize}
In short, while BigQuery’s partitioned tables implicitly move beyond the classic relational model for performance, MRA formalizes such structures in a principled way, providing both a general data model and a rich set of composable operations to the end users for manipulating them explicitly.

\subsection{DynamoDB}

DynamoDB is a popular NoSQL database, featuring key-value style data modeling. The key properties of DynamoDB data modelings are:
\begin{itemize}
\item Keys are modeled as two attributes, one called a partition key, one called a sort key. The partition key is used to partition the data by the key values, and for data of the same partition key, they are stored in the sorted order of the sort key.

\item Data stored within each partition are called items, where different items can have different attributes.
\end{itemize}

This data modeling can be captured by a slice relation with no loss of structure:
\begin{itemize}
\item The keys can be modeled as region schema with a partition attribute and a sort attribute. Regions partition the data into slice tuples -- which are partitions in the DynamoDB. The two-levels of storage due to the sort key can be viewed as storing regions in a 2-level way, where we group slice tuples of the same partition attribute together, and sort by the sort attribute.

\item The schema flexibility of the items is captured by the schema flexibility of feature schemas in a slice relation.
\end{itemize}

\end{document}